\documentclass[reprint,nofootinbib,superscriptaddress,floatfix]{revtex4-1} 
\usepackage{subcaption}
\usepackage{amsmath,amssymb}
\usepackage{color}
\usepackage{float}
\usepackage[normalem]{ulem}
\usepackage{cancel}
\usepackage{units}
\usepackage[Option]{overpic} 
\usepackage{tikz}
\usepackage{mathtools}
\usetikzlibrary{positioning,arrows}
\tikzset{
  arrow/.style={-latex, shorten >=1ex, shorten <=1ex}}

\def\C{\mathbb{C}}

\def\N{\mathbb{N}}
\def\Z{\mathbb{Z}}

\def\i{\mathrm {i}}
\def\e{\mathrm {e}}

\renewcommand{\Re}{\operatorname{Re}}
\renewcommand{\Im}{\operatorname{Im}}

\begin{document}

\title{Bandwidth and Conversion-Efficiency Analysis of Kerr Soliton Combs \\ in Dual-Pumped Resonators with Anomalous Dispersion}
\author{E.~Gasmi}
\affiliation{Institute for Analysis (IANA), Karlsruhe Institute of Technology, 76131 Karlsruhe, Germany}

\author{H.~Peng}
\affiliation{Institute of Photonics and Quantum Electronics (IPQ), Karlsruhe Institute of Technology, 76131 Karlsruhe, Germany}


\author{C.~Koos}
\affiliation{Institute of Photonics and Quantum Electronics (IPQ), Karlsruhe Institute of Technology, 76131 Karlsruhe, Germany}


\author{W.~Reichel}
\email[]{wolfgang.reichel@kit.edu}
\affiliation{Institute for Analysis (IANA), Karlsruhe Institute of Technology, 76131 Karlsruhe, Germany}

\begin{abstract} 
\noindent Kerr frequency combs generated in high-Q microresonators offer an immense potential in many applications, and predicting and quantifying their behavior, performance and stability is key to systematic device design. Based on an extension of the Lugiato-Lefever equation we investigate in this paper the perspectives of changing the pump scheme from the well-understood monochromatic pump to a dual-tone configuration simultaneously pumping two modes. For the case of anomalous dispersion we give a detailed study of the optimal choices of detuning offsets and division of total pump power between the two modes in order to optimize single-soliton comb states with respect to performance metrics like power conversion efficiency and bandwidth. Our approach allows also to quantify the performance metrics of the optimal single-soliton comb states and determine their trends over a wide range of technically relevant parameters.
\end{abstract}


\pacs{TODO}

\maketitle

\section{Introduction and main results} 

\noindent Optical frequency combs have revolutionized many applications, comprising optical frequency metrology \cite{udem2002optical}, spectroscopy \cite{picque2019frequency,yang2017microresonator}, optical frequency synthesizer \cite{jones2000carrier,spencer2018optical}, optical atomic clocks \cite{newman2019architecture}, ultrafast optical ranging \cite{trocha2018ultrafast}, and high-capacity optical communications using massively parallel wavelength-division multiplexing (WDM) \cite{marin2017microresonator}. The recent and rapid development of chip-scale Kerr soliton comb generators offers the prospects of realizing highly integrated devices which offer compactness, portability, and robustness, while being amenable to mass production and featuring low power consumption \cite{kippenberg2011microresonator}. Whereas Kerr soliton combs have conventionally been generated by using a monochromatic pump, dual-tone pumping configurations permit to achieve threshold-less comb generation in both normal and anomalous dispersion regimes \cite{PhysRevA.90.013811,lobanov2015generation}, while stabilizing the comb-tone spacing to a well-defined frequency \cite{papp2013parametric,strekalov2009generation}. The dual mode pumping scheme can be implemented either by using a phase- or intensity-modulated continuous-wave laser or two lasers with different wavelengths. Prior works theoretically investigated the dynamical properties of dissipative cavity soliton generation in a dual-mode-pumped Kerr microresonator by using the Lugiato-Lefever equation (LLE) with the addition of a secondary pump term \cite{weng2020formation}. However, a comprehensive study of the optimal pumping conditions for attaining the broadest comb bandwidth and the highest power conversion efficiency in the anomalous dispersion regime is still lacking.

In this paper we study a variant of the LLE based on a modification for dual-tone pumping \cite{Taheri_2017}, and we use this equation for a more detailed study of the benefits of dual-tone pumping. Focussing on resonators with anomalous dispersion, we find that dual-tone pumping allows to significantly improve key performance metrics of Kerr frequency combs such as bandwidth and power conversion efficiency. Mathematically, Kerr comb dynamics with a single pumped mode have been described by the LLE, a damped, driven and detuned nonlinear Schr\"odinger equation \cite{Lugiato_Lefever1987,Godey_et_al2014,Parra-Rivas2018}. Our modification of the LLE arises due to a forcing term which describes the pumping of two resonator modes instead of only a single one. 

\medskip

Using this equation as a base, we exploit numerical path continuation methods for a more detailed analysis of comb properties, the results of which can be summarized as follows: 
\begin{itemize}
\item[(1)] We show that pumping two modes is advantageous to pumping only one mode.
\item[(2)] We present heuristic insights for finding the optimal detuning parameters that provide the most localized single-soliton states. 
\item[(3)] We determined the optimal power distribution between the two pumped modes, which corresponds to a symmetric distribution where 50\% of the power is pumped into each mode\footnote{For purposes of simplifying the analysis this was exactly the case discussed by the authors in \cite{PhysRevA.90.013811}. Our findings validate their assumption of the pumps having equal amplitude and phase detuning.}. This power distribution simultaneously optimizes all performance metrics (comb bandwidth, full-width at half-maximum in time domain, and power conversion efficiency) in case equal detuning offsets between pump tones and nearest resonant modes are used.
\item[(4)] Under optimal power distribution we determined trends of the performance metrics w.r.t. varying dispersion and normalized total pump power.
\end{itemize}

\medskip

This paper is organized as follows: In Section~\ref{Lugiato-Lefever model for a dual-pumped ring resonator} we
introduce the Lugiato-Lefever model for a dual-pumped ring resonator. In Section~\ref{Heuristic} we present the main ideas for finding localized solitons in the case of pumping two adjacent modes. Section~\ref{OPD} is dedicated to the determination of the optimal power distribution between the two pumped modes. Here we use the comb bandwidth, the power conversion efficiency and the full-width at half-maximum as performance metrics. In Section~\ref{Trends} we provide trends for varying dispersion/forcing of this performance metrics under the provision of optimal equal power distribution between the two pumped modes. In Section~\ref{arbitrary_k1} we describe the optimal solitons achieved by pumping two arbitrarily distanced modes. Appendix A is dedicated to the derivation of the Lugiato-Lefever model for a dual-pumped ring resonator. In Appendix B we explain the details of the heuristic algorithm for finding localized solitons in the case of pumping two adjacent modes and Appendix C contains the heuristic for the case of pumping two arbitrarily distanced modes.

\section{Lugiato-Lefever model for a dual-pumped ring resonator}\label{Lugiato-Lefever model for a dual-pumped ring resonator}

\noindent Kerr comb dynamics are described by the LLE, a damped, driven and detuned nonlinear Schr\"odinger equation \cite{Lugiato_Lefever1987,Godey_et_al2014,Parra-Rivas2018}. As in \cite{Taheri_2017} we use a variant of the LLE modified for two-mode pumping, for which we provide a derivation of equation \eqref{PDE} starting from a system of nonlinear coupled mode equations in physical quantities in Appendix A. Using dimensionless, normalized quantities, this equation takes the form
\begin{equation}\label{PDE}
\i \frac{\partial a}{\partial \tau} =-d a''-(\i-\zeta_0)a-|a|^2a+\i f_0+\i f_1\e^{\i(k_1x-\nu_1 \tau)}.
\end{equation}
Here, $a(\tau,x)$ is $2\pi$-periodic in $x$ and represents the optical intracavity field as a function of normalized time $\tau=\kappa t/2$ and angular position $x\in[0,2\pi]$ within the ring resonator. The constant $\kappa>0$ describes the cavity decay rate and $d=2d_2/\kappa>0$ quantifies the anomalous dispersion in the system ($2d_2$ corresponds to the difference between two neighboring FSRs at the center frequency $\omega_0$). Since the numbering $k\in\Z$ of the resonant modes in the cavity is relative to the first pumped mode $k_0 = 0$ we denote with $k_1 \in\N$ the second pumped mode (there is no loss of generality to take $k_1$ as a positive integer since $k_1$ and $-k_1$ are symmetric modes). Since there are now two pumped modes there will also be two normalized detuning parameters denoted by $\zeta_0=2(\omega_0-\omega_{p_0})/\kappa$ and $\zeta_1=2(\omega_{k_1}-\omega_{p_1})/\kappa$. They describe the offsets of the input pump frequencies $\omega_{p_0}$ and $\omega_{p_1}$ to the closest resonance frequency $\omega_0$ and $\omega_{k_1}$ of the microresonator, respectively. Finally $f_0,f_1$ represent the normalized power of the input pumps. If we set $\Delta \zeta$=$\zeta_0-\zeta_1$ and $\nu_1=\Delta \zeta+dk_1^2$ then (after several transformations, cf. Appendix A) equation \eqref{PDE} emerges with the specific form of the second pump $f_1\e^{\i(k_1x-\nu_1 \tau)}$.

In the case $f_1=0$, equation \eqref{PDE} amounts to the case of pumping only one mode. This case has been thoroughly studied, e.g. in \cite{GaertnerTrochaMandel2018_1000089036,Godey_et_al2014,Godey_2017,Parra-Rivas2014,Parra-Rivas2016,Parra-Rivas2018,mandel_reichel,Miyaji_Ohnishi_Tsutsumi2010,DelHara_periodic,Perinet}. In this paper we are interested in the case $f_1\neq 0$. The particular form of the pump term $\i f_0+\i f_1\e^{\i(k_1x-\nu_1 \tau)}$ suggests to perform a change of variables into a moving coordinate $s=x-\omega \tau$ with $\omega=\nu_1/k_1$ and study solutions of \eqref{PDE} of the form $a(\tau,x)=u(x-\omega \tau)$. These traveling-wave solutions propagate with speed $\omega$ in the resonator, and their profile $u$ solves the stationary ordinary differential equation 
\begin{equation}\label{TWE}
-d u'' +\i \omega u'-(\i-\zeta_0)u-|u|^2 u+\i f_0 +\i f_1 \e^{\i k_1 s}=0,
\end{equation}
where $u$ is again $2\pi$-periodic in $s$. 
In Fourier modes $a$ and $u$ are represented as $a(\tau,x)=\sum_{k\in\Z} \hat a_k(\tau) \e^{\i k x}$, $u(s)=\sum_{k\in \Z}\hat u_k \e^{\i k s}$. The intracavity power $P$ of the field $a$ at time $\tau$ is given by 
\begin{equation*}
P=\sum_{k\in\Z} |\hat a_k(\tau)|^2=\frac{1}{2\pi}\int_0^{2\pi} \, |a(\tau,x)|^2 \, dx.
\end{equation*}
Since the Fourier modes of $a$ and $u$ are related by $\hat a_k(\tau)=\hat u_k  \e^{-\i k \omega \tau}$ one finds $P=\sum_{k\in\Z} |\hat u_k|^2=\frac{1}{2\pi}\int_0^{2\pi} \, |u(s)|^2 \, ds$. In particular, $P$ is independent\footnote{In fact, the power $|\hat u_k|^2=|\hat a_k(\tau)|^2$ in each mode is independent of time.} of the time, and since $\int_0^{2\pi}|u|^2\,ds = \Re\int_0^{2\pi} (f_0+f_1\e^{\i k_1s})\bar u\,ds$ we see that $P\leq f^2 \coloneqq f_0^2+f_1^2$, i.e., the intracavity power cannot exceed the normalized total input power. Details are given at the end of Appendix A. Here, the notation $\bar z$ denotes the complex conjugate of the complex number $z\in \C$.

\section{Heuristic for finding localized solitons in the case of pumping two adjacent modes} \label{Heuristic} \noindent 
In the following section, we explain the main idea of the heuristic for finding strongly localized solutions of \eqref{TWE}, where two adjacent modes are pumped, i.e. the pumped modes are $k_0=0$ and $k_1=1$. In Appendix B we provide a more detailed explanation, and in Appendix C we show how the heuristic can be adapted to arbitrary values of $k_1 \in \N$. The parameters $d>0$, $k_1=1$, $f_0$ and $f_1$ are fixed, and our goal is to find optimally localized solutions by varying the parameters $\zeta_0$ and $\omega$ since they can be influenced by the choice of the pump frequencies $\omega_{p_0}$ and $\omega_{p_1}$ through the relations 
$$
\zeta_0 = \frac{2}{\kappa}\bigl(\omega_0-\omega_{p_0}\bigr), \, \omega = \frac{2}{\kappa}\bigl(\omega_0-\omega_{p_0}-(\omega_1-\omega_{p_1})+d_2\bigr).
$$
Optimality is understood as minimality with respect to the full-width at half-maximum (FWHM) of the field distribution $|u|^2$ in the time domain. We have developed our heuristic by using the Matlab package \texttt{pde2path} (cf. \cite{UEC14}, \cite{DOH14}) which has been designed to numerically treat continuation and bifurcation in boundary value problems for systems of PDEs.\footnote{Continuation and bifurcation solvers for boundary value problems (on which \texttt{pde2path} is based) allow to globally study the variety of different stationary comb states by exploiting the full range of technically available parameters. In contrast, time-integration solvers mostly only allow to access specific comb states which strongly depend on the chosen device parameters and initial conditions.} 

\medskip

In short, the basic algorithm is explained as follows: First we obtain a single-peak solution for the correct value of the parameter $f_1$ (ignoring the values of the parameters $\zeta_0$ and $\omega$). Then we alternately run a continuation algorithm by varying either the $\zeta_0$- or the $\omega$-parameter (while keeping the other parameter fixed) and detect among the continued solutions the soliton $u$ with minimal FWHM of $|u|^2$ in the time domain. We denote the soliton obtained from the $j$-th $\zeta_0$-optimization as $A_j$ and the one obtained from the $j$-th $\omega$-optimization as $B_j$. We stop the algorithm when the relative change of the FWHM of $B_{j+1}$ and $B_j$ is sufficiently small. In our numerical experiments it was always sufficient to perform at most three optimizations in both of the variables $\zeta_0$ and $\omega$.  

\begin{figure*}[t]
\centering
\begin{minipage}[t]{0.3\textwidth}
\includegraphics[width=\columnwidth]{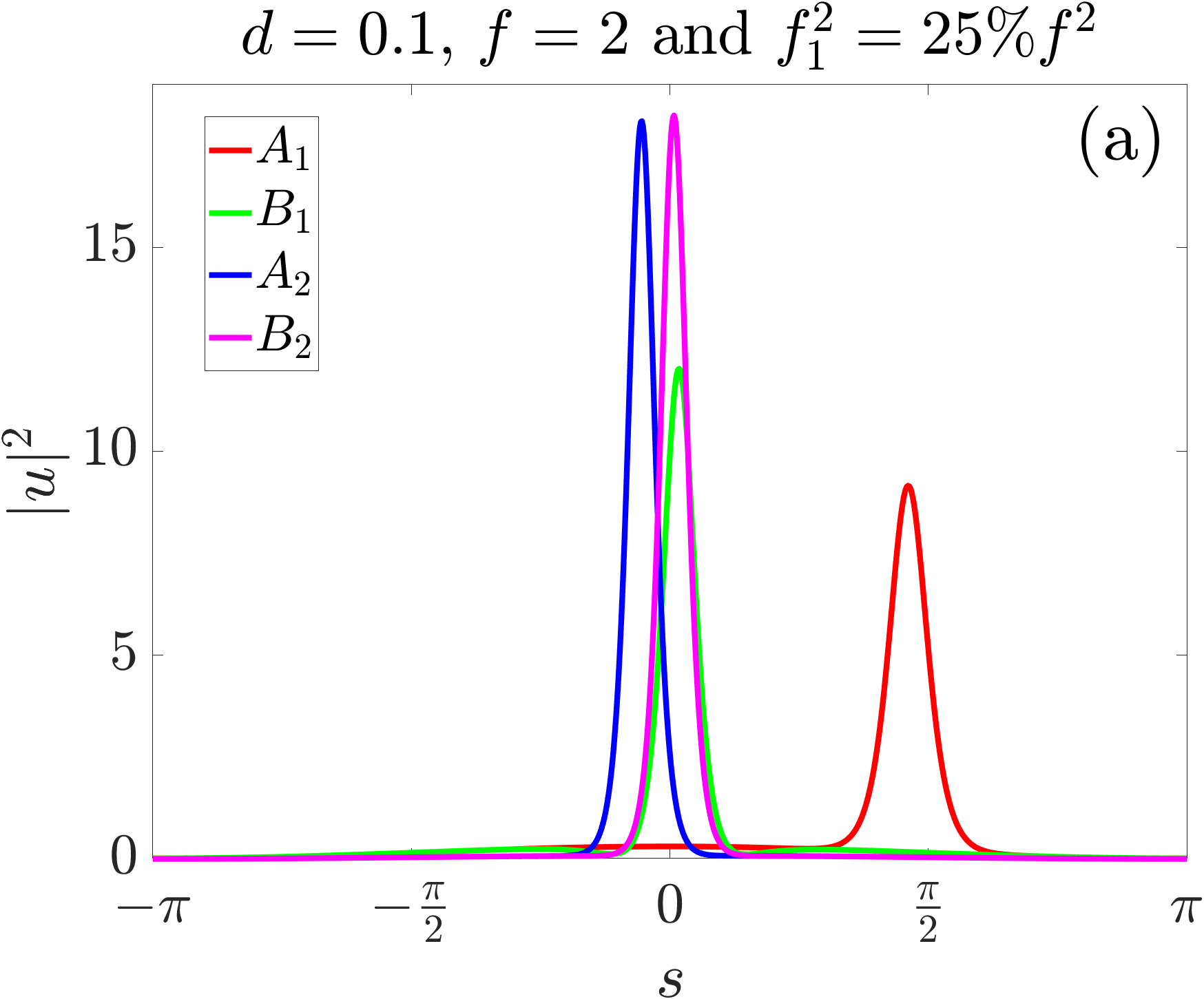} \\[0.5cm]
\includegraphics[width=\columnwidth]{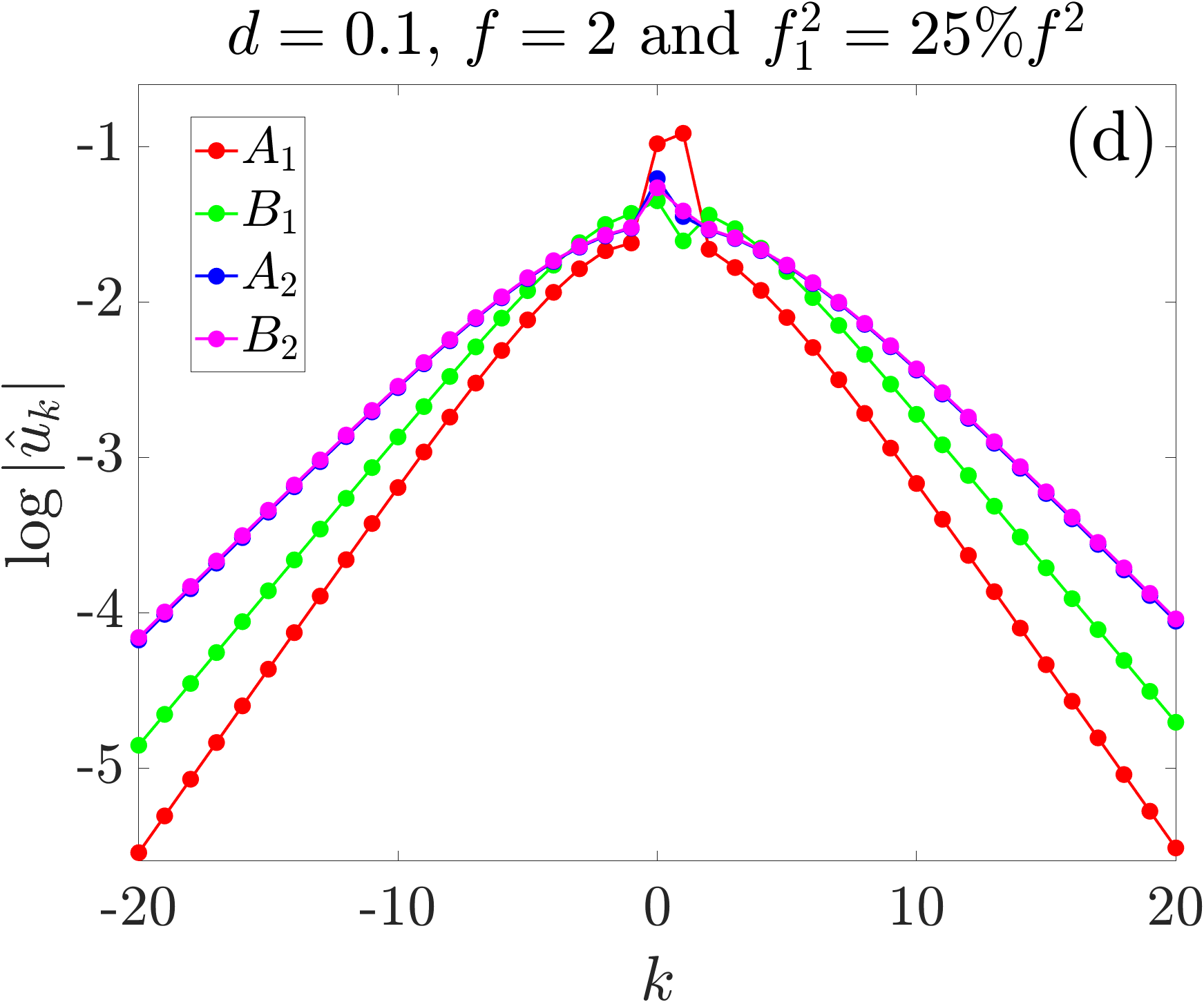} \\[0.5cm]
\includegraphics[width=\columnwidth]{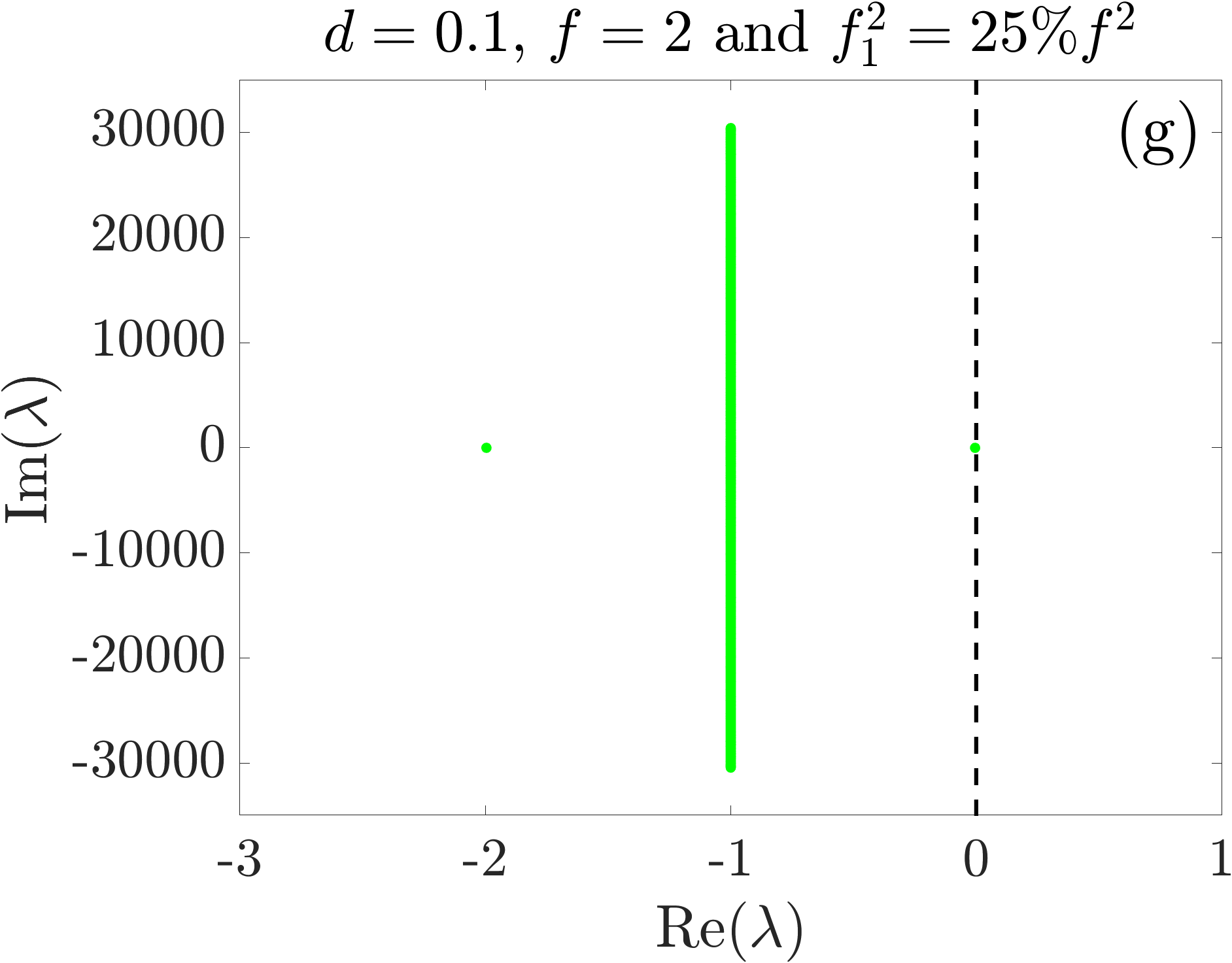}
\end{minipage} \hspace{0.5cm}
\begin{minipage}[t]{0.3\textwidth}
\includegraphics[width=\columnwidth]{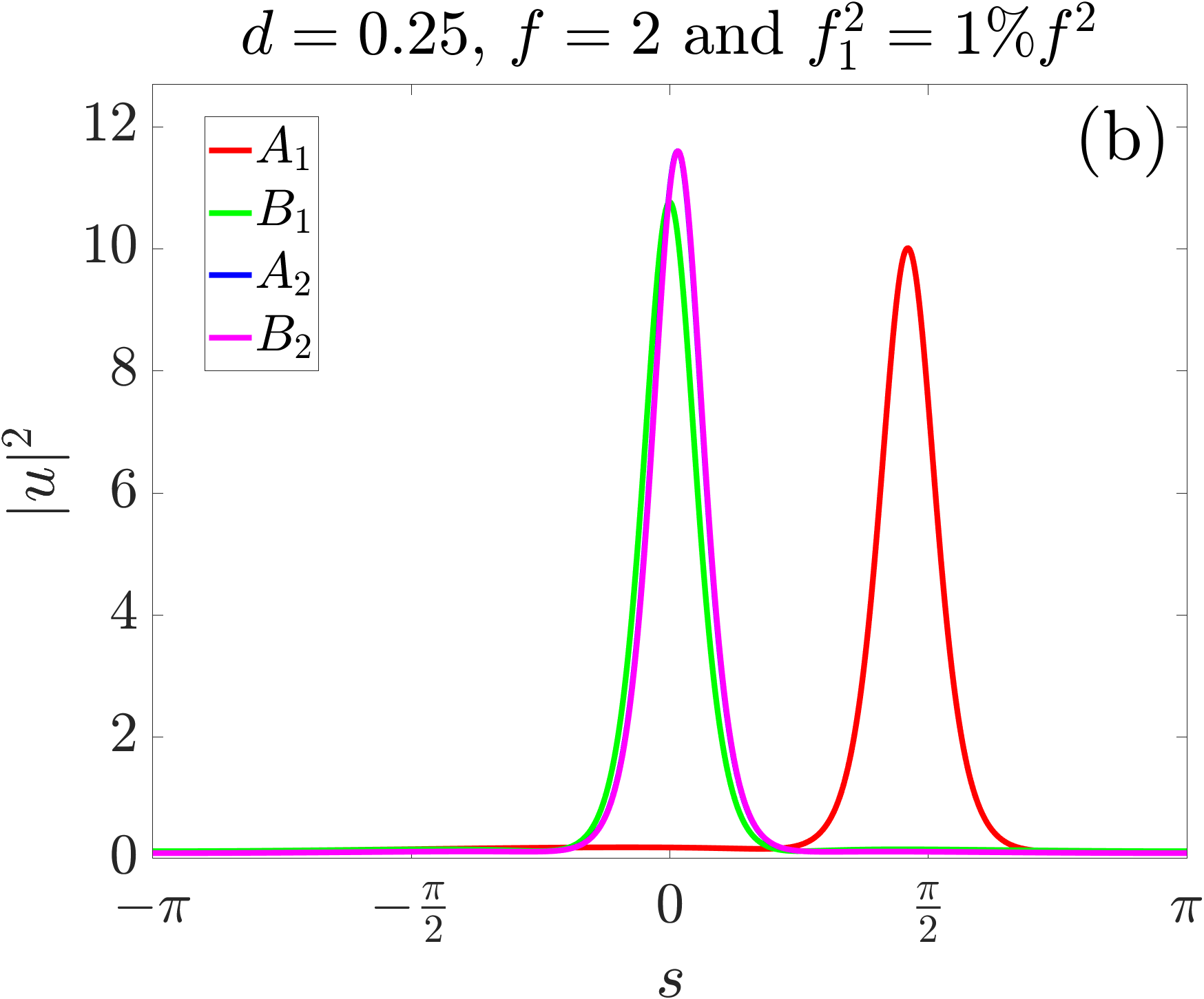} \\[0.5cm]
\includegraphics[width=\columnwidth]{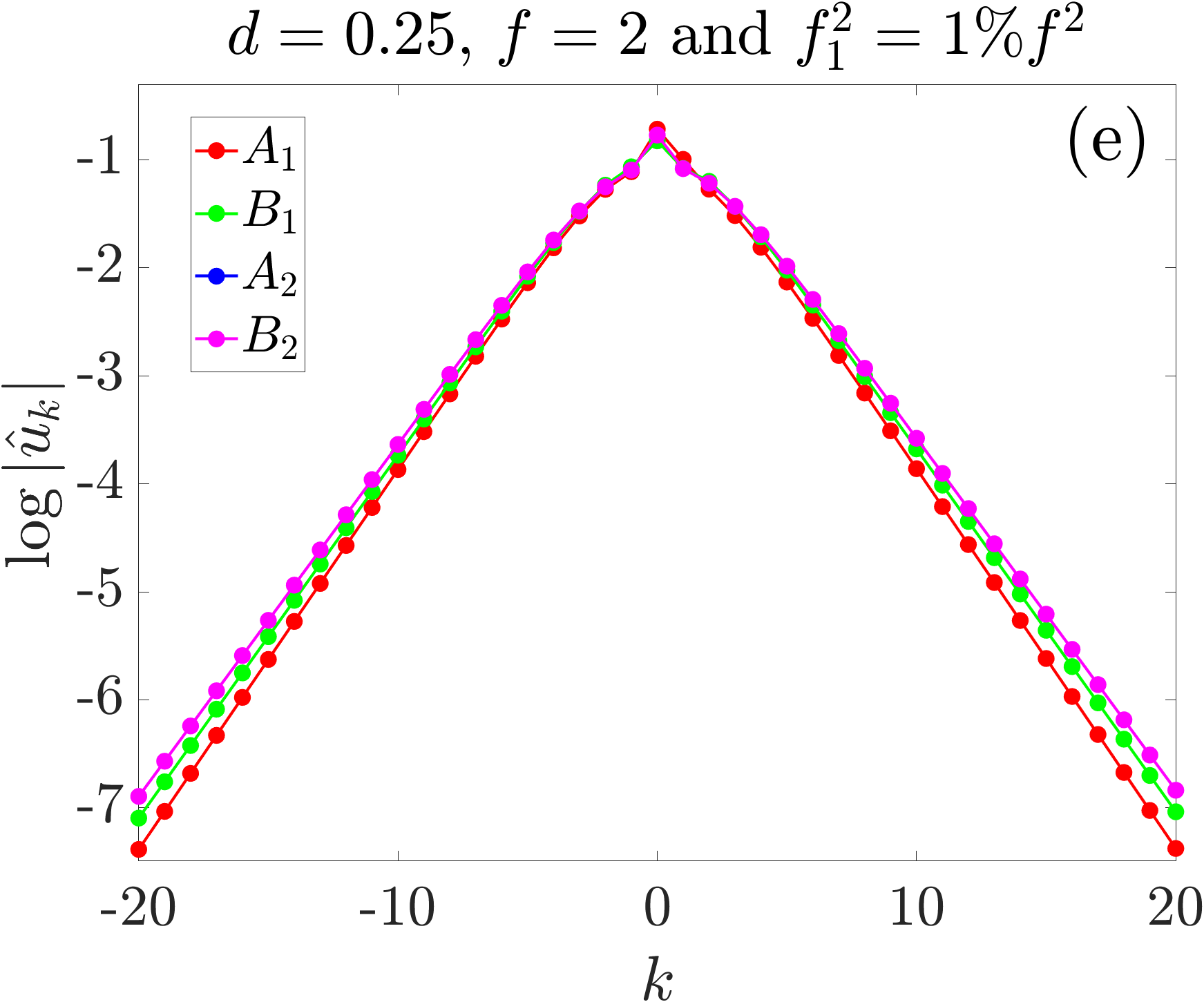} \\[0.5cm]
\includegraphics[width=\columnwidth]{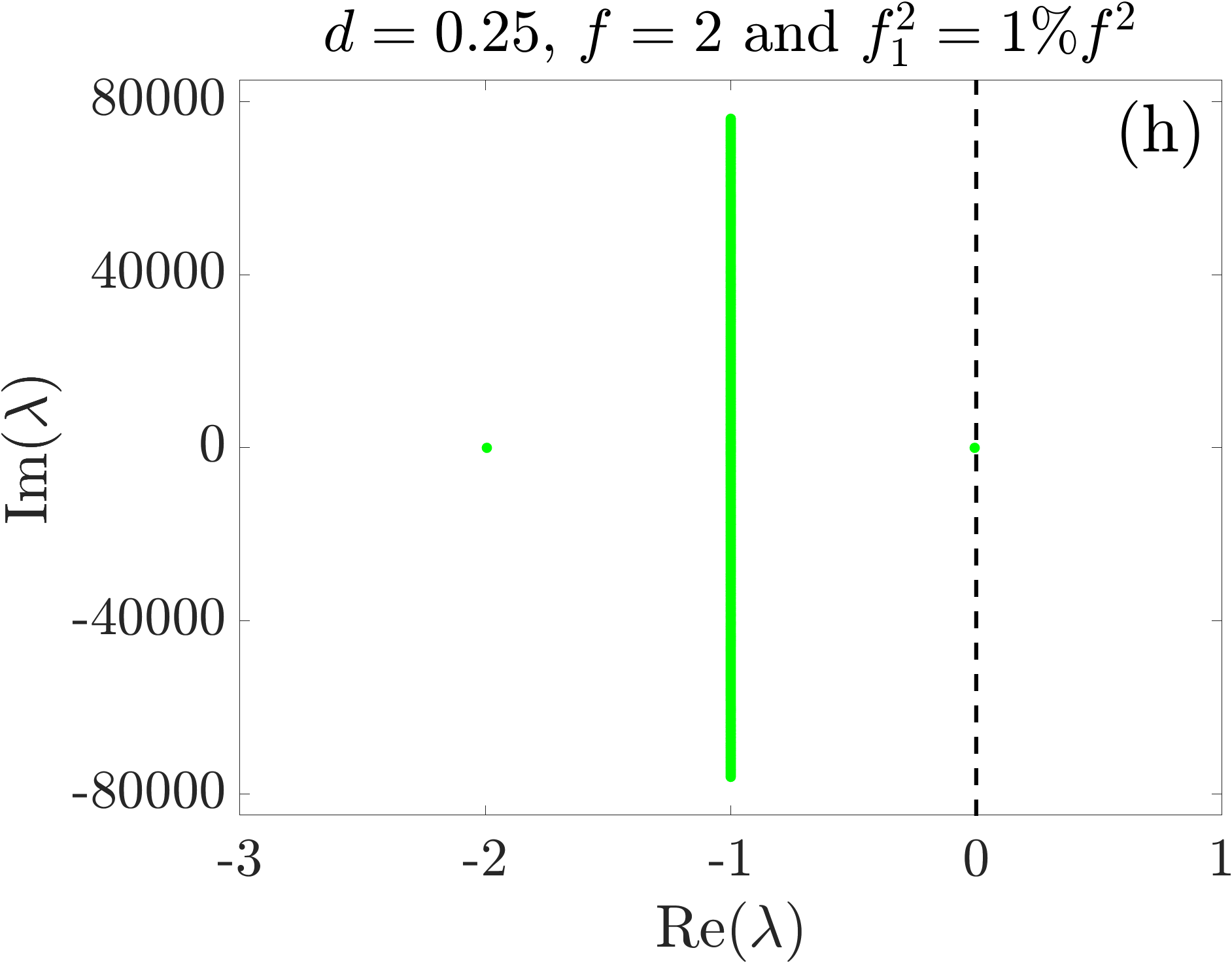}
\end{minipage} \hspace{0.5cm}
\begin{minipage}[t]{0.3\textwidth}
\includegraphics[width=\columnwidth]{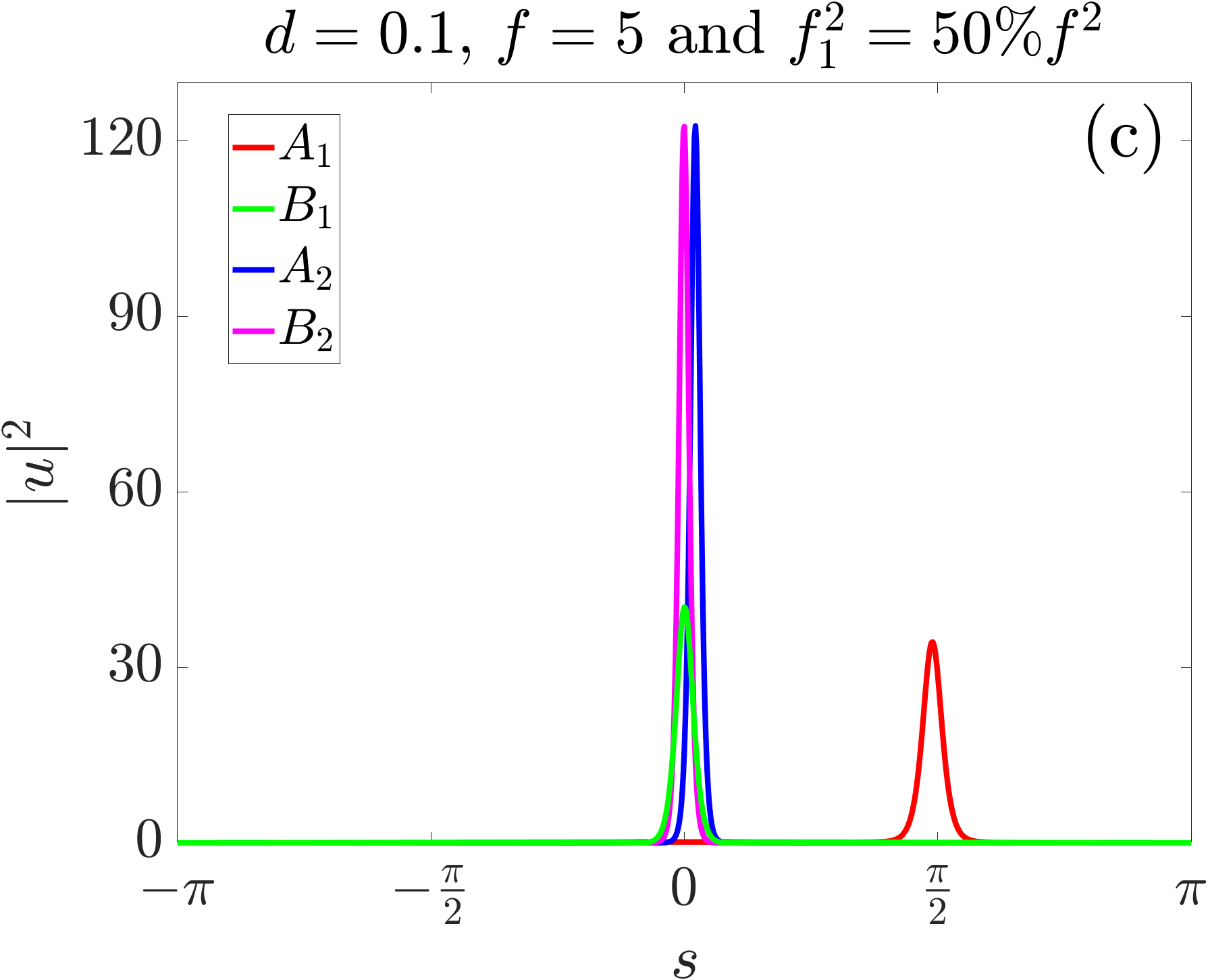} \\[0.5cm]
\includegraphics[width=\columnwidth]{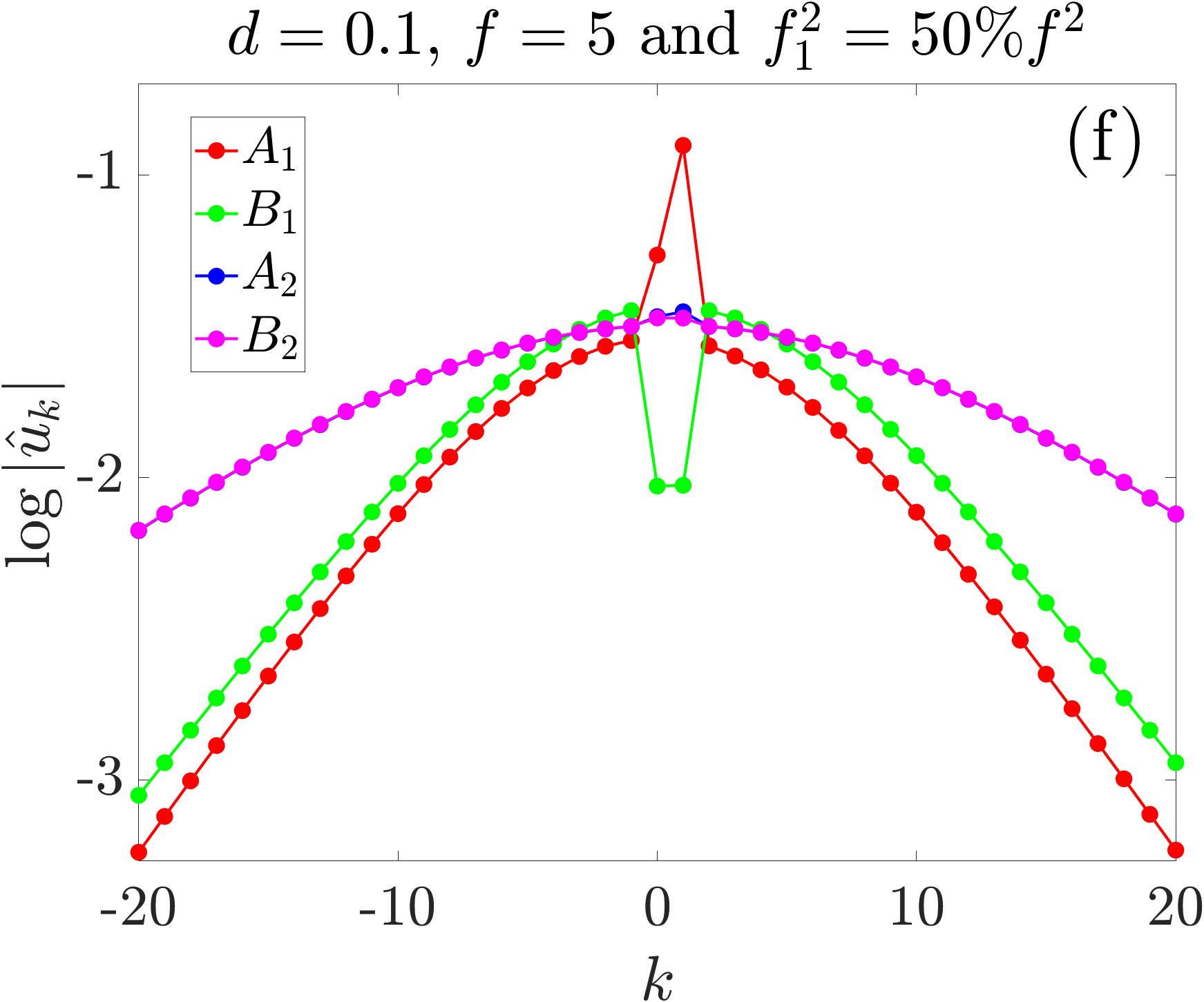} \\[0.5cm]
\includegraphics[width=\columnwidth]{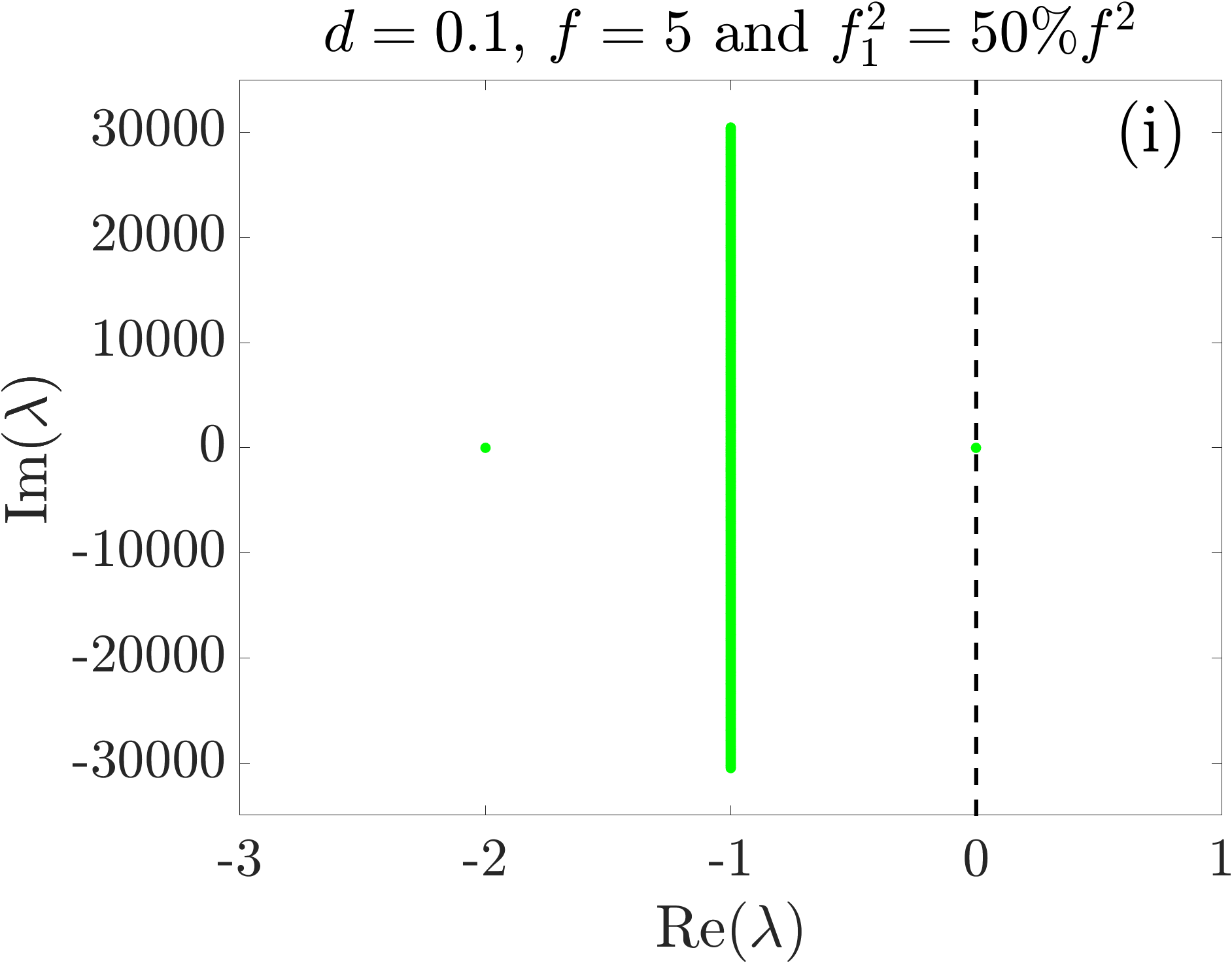}
\end{minipage}
\caption{Spatial and spectral power distributions of the solitons obtained from two iterations (two $\zeta_0$-steps leading to $A_1$, $A_2$ and two $\omega$-steps leading to $B_1$, $B_2$) and a stability plot for $B_3$ (obtained from the third $\omega$-step) for three different choices of the parameters $d$, $f$ and $f_1$. Every column corresponds to one choice. In (g)-(i) we plotted in green the spectrum of the finite-element discretization of the linearized operator $L$ at $B_3$. The black dashed line in (g)-(i) represents the imaginary axis. The spectrum lies to the left of the imaginary axis so that the solitons are spectrally stable.}
\label{fig:OptimizationSolitonsCombs}
\end{figure*}

\medskip

In Fig.~\ref{fig:OptimizationSolitonsCombs}(a)-(c) we plotted the spatial power distributions of the solitons $A_j$ and $B_j$ for two iteration steps $j=1,2$ and three different choices of the parameters $d$, $f$ and $f_1$. It is well visible that the solitons get more localized after every optimization step and that the solitons $A_2$ and $B_2$ from the second iteration steps do not differ significantly. In the second column of Fig.~\ref{fig:OptimizationSolitonsCombs} in (b) and (e) the blue soliton $A_2$ is not visible, since it is covered by the almost identical magenta soliton $B_2$. In the second row Fig.~\ref{fig:OptimizationSolitonsCombs}(d)-(f) we show the spectral power distributions. The final magenta comb $B_2$ covers almost entirely the blue comb $A_2$. The third row of Fig.~\ref{fig:OptimizationSolitonsCombs} contains information on the spectral stability of the optimized solitons. This will be explained next.

\medskip

\textit{Stability of optimal solitons.} To investigate the stability of the solitons, we use the transformation $a(\tau,x)=b(\tau,x-\omega \tau)$ to rewrite \eqref{PDE} as
\begin{equation}\label{TWPDE}
\frac{\partial b}{\partial \tau} = -\i\bigl(-d b'' +\i \omega b'-(\i-\zeta_0)b-|b|^2 b+\i f_0 +\i f_1 \e^{\i k_1 s}\bigr),
\end{equation}
where $b$ is again $2\pi$-periodic in $s$. Solutions $u$ of \eqref{TWE} correspond to stationary solutions $b(\tau,s)=u(s)$ of \eqref{TWPDE}. Spectral stability is based on the following considerations. Let $b(\tau,s)\approx u(s)+\phi(s)\e^{\lambda \tau}+\psi(s)\e^{\bar\lambda \tau}$ with $2\pi$-periodic functions $\phi, \psi$, and insert this ansatz into \eqref{TWPDE}. After keeping only the linear terms in $\phi$ and $\psi$, we find that $\phi, \psi$ have to satisfy the eigenvalue equation 
$$
L \begin{pmatrix} \phi \\ \bar\psi \end{pmatrix} = \lambda \begin{pmatrix} \phi \\ \bar\psi \end{pmatrix}
$$
with the linearized operator  
\begin{widetext}
$$
L = \begin{pmatrix} 
\i d \frac{d^2}{ds^2} + \omega \frac{d}{ds} -1-\i \zeta+2\i |u|^2 & \i u^2 \\
-\i \bar u^2 & -\i d \frac{d^2}{ds^2} + \omega \frac{d}{ds} -1+\i \zeta-2\i |u|^2
\end{pmatrix}.
$$
\end{widetext}
We see that the perturbation $\phi(s)\e^{\lambda \tau}+\psi(s)\e^{\bar\lambda \tau}$ will tend to zero if and only if the eigenvalues $\lambda$ of $L$ lie in the left complex plane. Using this criterion, we found that the optimized solitons (optimized w.r.t. $\zeta_0$ and $\omega$ by the above heuristic) discussed in this section are all spectrally stable. To show this, we computed the eigenvalues of the finite-element discretization of the operator $L$ and observed that they entirely belong to the left complex plane, cf. Fig.~\ref{fig:OptimizationSolitonsCombs}(g)-(i). One sees that there is always an eigenvalue very close to $0$. The reason for this is the following. The optimized solitons are found near turning points along branches of the $\zeta_0$-continuation, cf. Appendix B. These turning points are necessarily associated with a $0$ eigenvalue of the linearized operator $L$. Hence, for $u$ being in the vicinity of a turning point, there will be an eigenvalue of $L$ very close to $0$.

\section{Optimal power distribution when pumping two adjacent modes}\label{OPD} \noindent 
In this section we answer the question which amount of the normalized total input power $f^2=f_0^2+f_1^2$ needs to be pumped into each mode in order to obtain the best soliton, i.e., we determine the optimal power distribution between the two pumped modes. The power distribution is described as $(f_0,f_1)=(f\cos\varphi,f\sin\varphi)$ with $\varphi \in [0,2\pi)$. As before, we assume anomalous dispersion $d>0$ and fix the indices $k_0=0$ and $k_1=1$ of the two pumped modes. Additionally, the normalized total input power $f^2$ is given. Armed with the heuristic from Section~\ref{Heuristic} we are able to identify for any fixed $\varphi\in [0,2\pi)$ a 1-soliton with the strongest spatial localization, i.e., with minimal FWHM. 

\medskip

Using this approach, we calculate for each such a comb state $u(s)=\sum_{k\in\Z} \hat u_k \e^{\i k s}$ the power conversion efficiency (PCE), the comb bandwidth (CBW) and its FWHM. The PCE is defined as the ratio $P_{\text{FC}}/f^2$ between intracavity comb power 
\begin{align*}
P_{\text{FC}}&=\sum_{k\in\Z\setminus\{0,1\}} |\hat u_k|^2 +\frac{f_1^2}{f^2}|\hat u_0|^2+\frac{f_0^2}{f^2}|\hat u_1|^2 \\
&=\sum_{k\in\Z\setminus\{0,1\}} |\hat u_k|^2 +\sin^2(\varphi)|\hat u_0|^2+\cos^2(\varphi)|\hat u_1|^2
\end{align*}
and the normalized total input power. Note that the intracavity comb power is a weighted sum over the power in each mode. The weights $f_j^2/f^2$, $j=0,1$ of the power of the zero mode and the first mode are such that $f_1=0$ or $f_0=0$ lead to the usual definition of PCE and $f_0\to\infty$ or $f_1\to \infty$ lead to an exclusion of the power contributed by the zero or first mode, respectively. The CBW is defined via the 3dB points, i.e.,
$$
\text{CBW} = k^*_l+k^*_r
$$
with minimal integers $k^*_l>0$ and $k^*_r>0$ which fulfill 
$$|\hat u_{-k^*_l}|^2 \leq \frac{1}{2}|\hat u_{-1}|^2, \quad
|\hat u_{1+k^*_r}|^2 \leq \frac{1}{2}|\hat u_2|^2,$$
respectively. Note that the 3dB comb bandwidth is defined with respect to the power $|\hat u_{-1}|^2$ and $|\hat u_2|^2$ of the modes directly adjacent to the pumped modes rather than the power $|\hat u_0|^2$ and $|\hat u_1|^2$ of the pumped modes themselves. 

\medskip

\begin{figure*}
\centering
\begin{minipage}[t]{0.305\textwidth}
\includegraphics[width=\columnwidth]{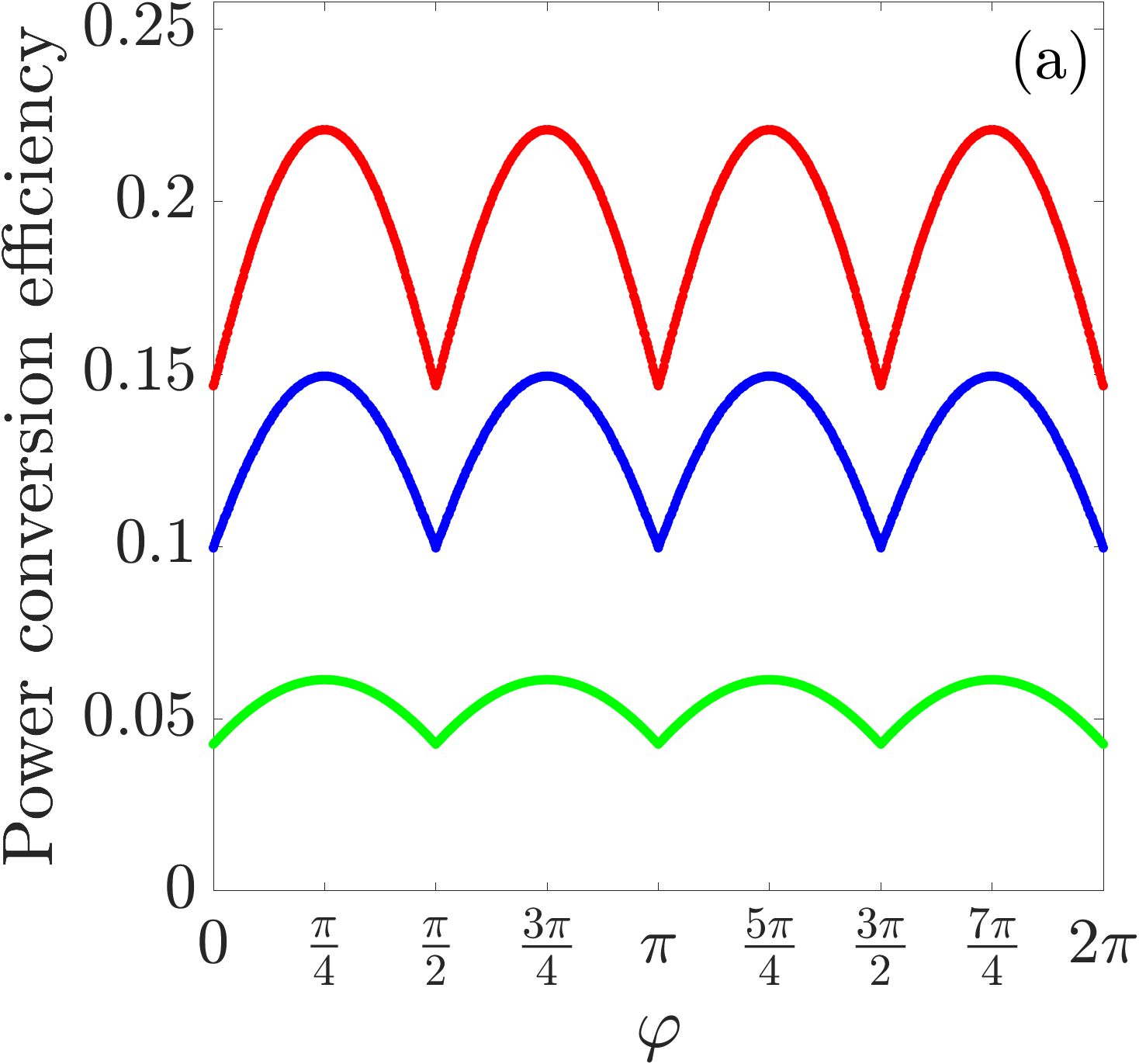}
\end{minipage} \hspace{0.5cm}
\begin{minipage}[t]{0.2905\textwidth}
\includegraphics[width=\columnwidth]{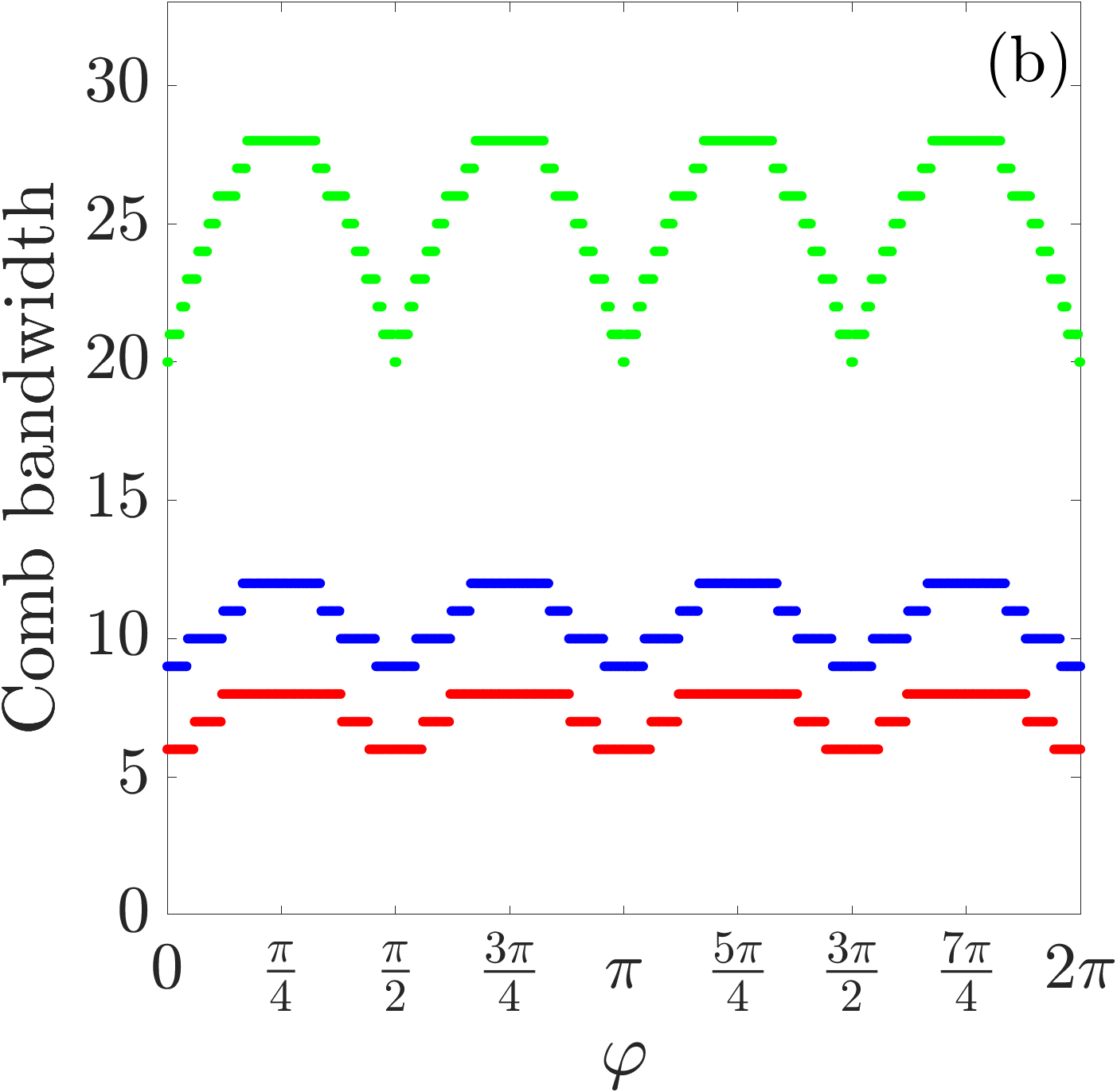}
\end{minipage} \hspace{0.5cm}
\begin{minipage}[t]{0.295\textwidth}
\includegraphics[width=\columnwidth]{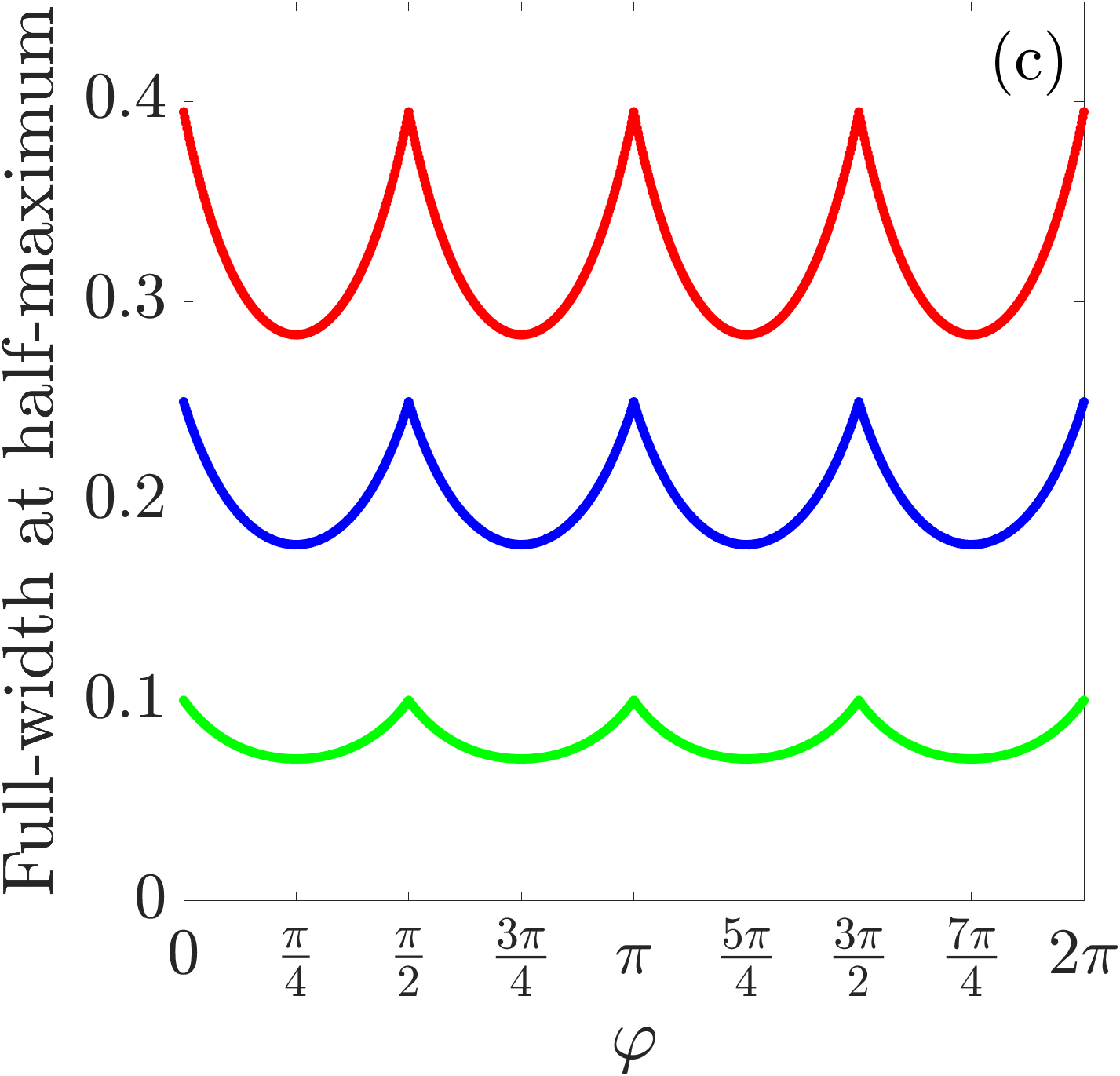}
\end{minipage}
\caption{Power conversion efficiency, comb bandwidth and full-width at half-maximum as a function of $\varphi$ for three different examples. The blue curves correspond to $d=0.1$ and $f=2$, the red ones to $d=0.25$ and $f=2$ as well as the green ones to $d=0.1$ and $f=5$.}
\label{fig:PerformanceMetricsUnderPowerDistribution}
\end{figure*}

To find the optimal power distribution between the zero mode and the first mode we performed a parameter study in $\varphi$ for three different examples, cf. Fig.~\ref{fig:PerformanceMetricsUnderPowerDistribution}. In the first example we chose $d=0.1$ and $f=2$, in the second example we kept $f=2$ but changed the dispersion to $d=0.25$ while in the last example we kept $d=0.1$ and changed the forcing to $f=5$. For these three examples we computed the most localized 1-soliton for $\varphi\in[0,2\pi)$ based on the heuristic of Section~\ref{Heuristic} and evaluated the PCE, the CBW as well as the FWHM of the resulting comb state. 

\medskip

The results depicted in Fig.~\ref{fig:PerformanceMetricsUnderPowerDistribution} clearly demonstrate the advantages of dual-tone pumping, in particular when using equal power in both modes. In all of the examples PCE and CBW increase while the FWHM decreases with $\varphi \in [0,\pi/4]$. Moreover, as we will explain at the end of this section, PCE, CBW and FWHM are $\pi/2$-periodic and symmetric w.r.t. $\pi/4$. We conclude that 
\begin{itemize}
\item[(i)] pumping two modes is advantageous to pumping only one mode,
\item[(ii)] PCE, CBW and FWHM are monotonic functions of $|f_0|+|f_1|=|f|(|\cos\varphi|+|\sin\varphi|)$,
\item[(iii)] the optimal case arises for equal pump powers $|f_0|=|f_1|$. 
\end{itemize}

\begin{figure*}
\centering
\begin{minipage}[t]{0.29\textwidth}
\includegraphics[width=\columnwidth]{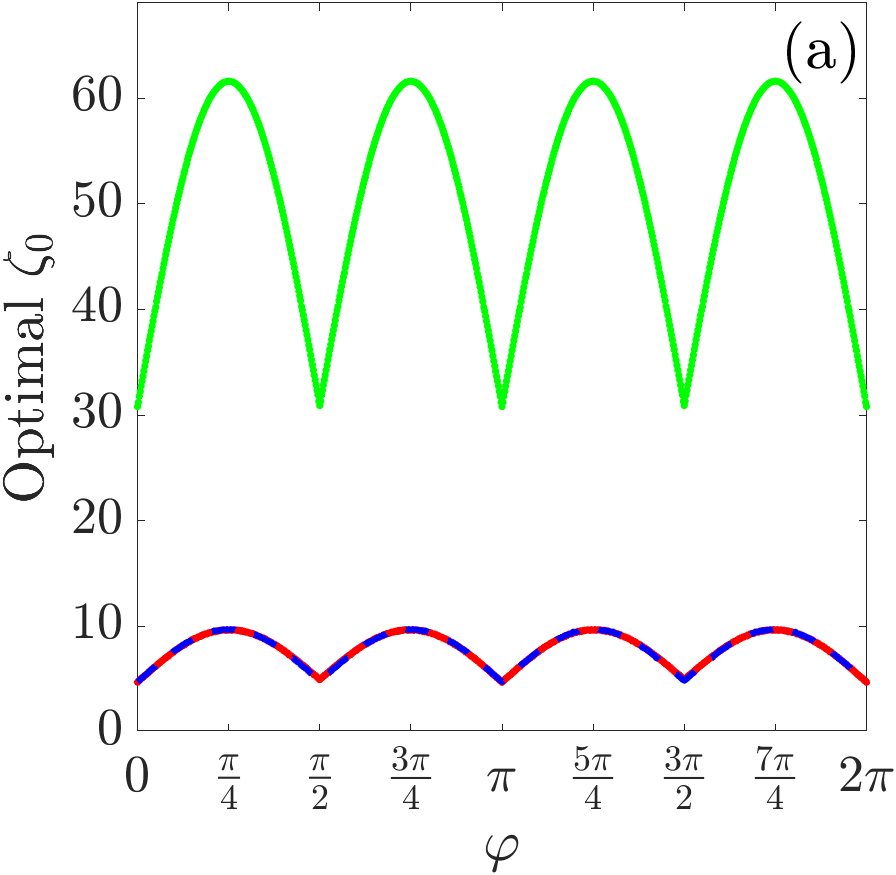}
\end{minipage} \hspace{0.5cm}
\begin{minipage}[t]{0.305\textwidth}
\includegraphics[width=\columnwidth]{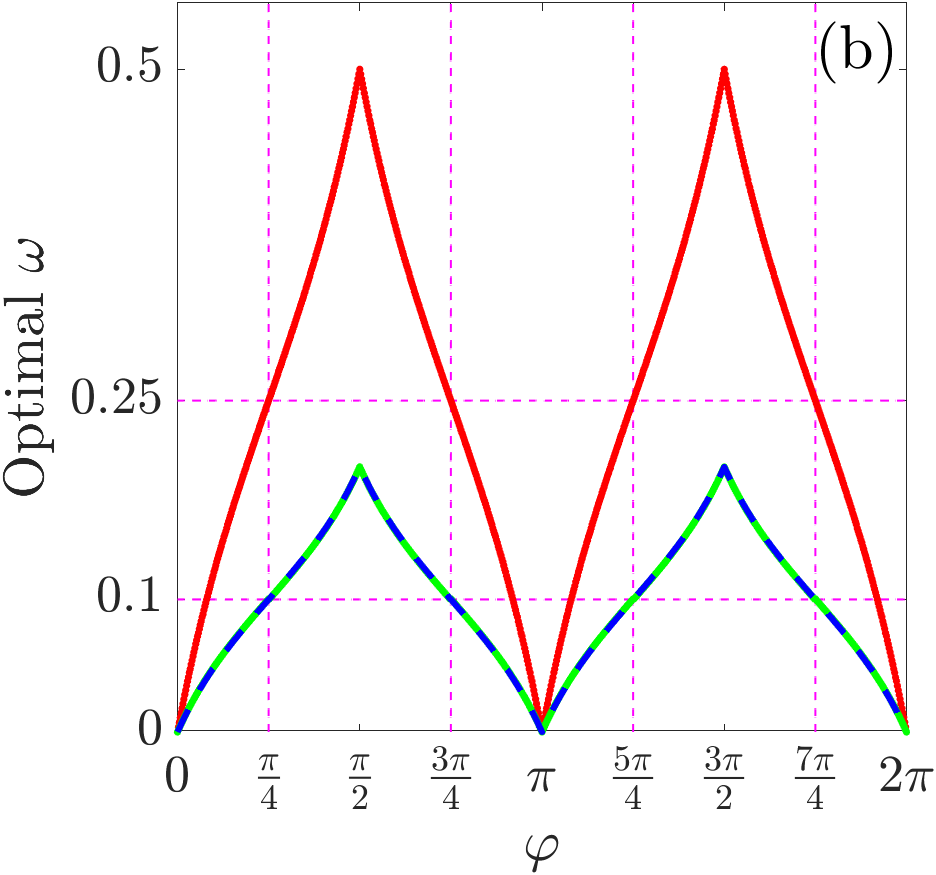}
\end{minipage} \hspace{0.5cm}
\begin{minipage}[t]{0.31\textwidth}
\includegraphics[width=\columnwidth]{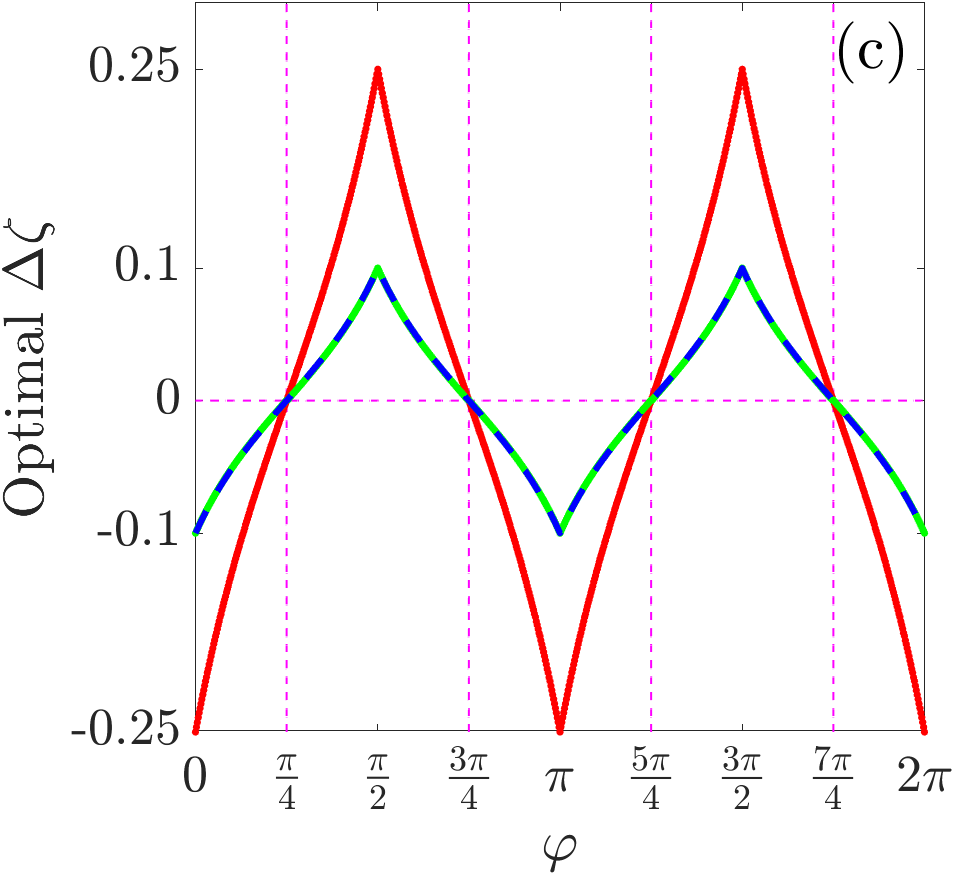}
\end{minipage} 
\caption{Optimal values of $\zeta_0$, $\omega$ and $\Delta \zeta$ as a function of $\varphi$ for three different examples. The blue curves correspond to $d=0.1$ and $f=2$, the red ones to $d=0.25$ and $f=2$ as well as the green ones to $d=0.1$ and $f=5$. The blue and the red curves in (a) as well as the blue and the green curves in (b) and (c) are plotted dashed so that one of the curves is not completely covered by the other one. The dashed lines colored in magenta in (b) emphasize that the optimal value of $\omega$ coincides with the dispersion
$d$ in the optimal case $|f_0| = |f_1|$ where $\varphi \in \{\pi/4,3\pi/4,5\pi/4,7\pi/4\}$. The dashed lines colored in magenta in (c) emphasize that the optimal value of $\Delta \zeta$ vanishes in the optimal case $|f_0| = |f_1|$ where $\varphi \in \{\pi/4,3\pi/4,5\pi/4,7\pi/4\}$.}
\label{fig:OptimalValuesZetaOmega}
\end{figure*}

In Fig.~\ref{fig:OptimalValuesZetaOmega}(a)-(b) we plotted the optimal values of $\zeta_0$ and $\omega$ (for which the most localized soliton was found) against $\varphi$. Since $k_1=1$ we have $\omega=\Delta\zeta+d$ so that the optimal value of $\omega$ can be easily translated into an optimal value of $\Delta \zeta$. We added in Fig.~\ref{fig:OptimalValuesZetaOmega}(c) a plot of the optimal value of $\Delta \zeta$ against $\varphi$ since the normalized detuning difference $\Delta \zeta=\zeta_0-\zeta_1$ is the physically more tangible quantity while from a mathematical point of view it is more convenient to work with $\omega$. In all of the examples the optimal values of $\zeta_0$, $\omega$ and $\Delta \zeta$ increase with $\varphi\in(0,\pi/4]$. Once again we observe several symmetries, which we will address in the end of this section. We further conclude that 
\begin{itemize}
\item[(iv)] the optimal value of $\zeta_0$ is almost independent of $d$,
\item[(v)] the optimal value of $\omega$ is almost independent of $f$,
\item[(vi)] the optimal value of $\omega$ coincides with the dispersion $d$ in case of optimal power distribution $|f_0|=|f_1|$.
\end{itemize} 
As $\omega=\Delta \zeta+d$, (vi) means $\Delta\zeta=0$, i.e., optimal solitons require equal detuning distances $\omega_0-\omega_{p_0}=\omega_1-\omega_{p_1}$ in case of equal power distribution $|f_0|=|f_1|$. From Fig.~\ref{fig:OptimalValuesZetaOmega}(c) we further find that the optimal values for $\zeta_0$ and $\zeta_1$ satisfy the relation $|f_0|>|f_1| \Leftrightarrow \zeta_0<\zeta_1$, i.e., pumping more power into one mode is compensated by a larger detuning for the second mode. 

\medskip

\begin{figure*}
\centering
\begin{minipage}[t]{0.3\textwidth}
\includegraphics[width=\columnwidth]{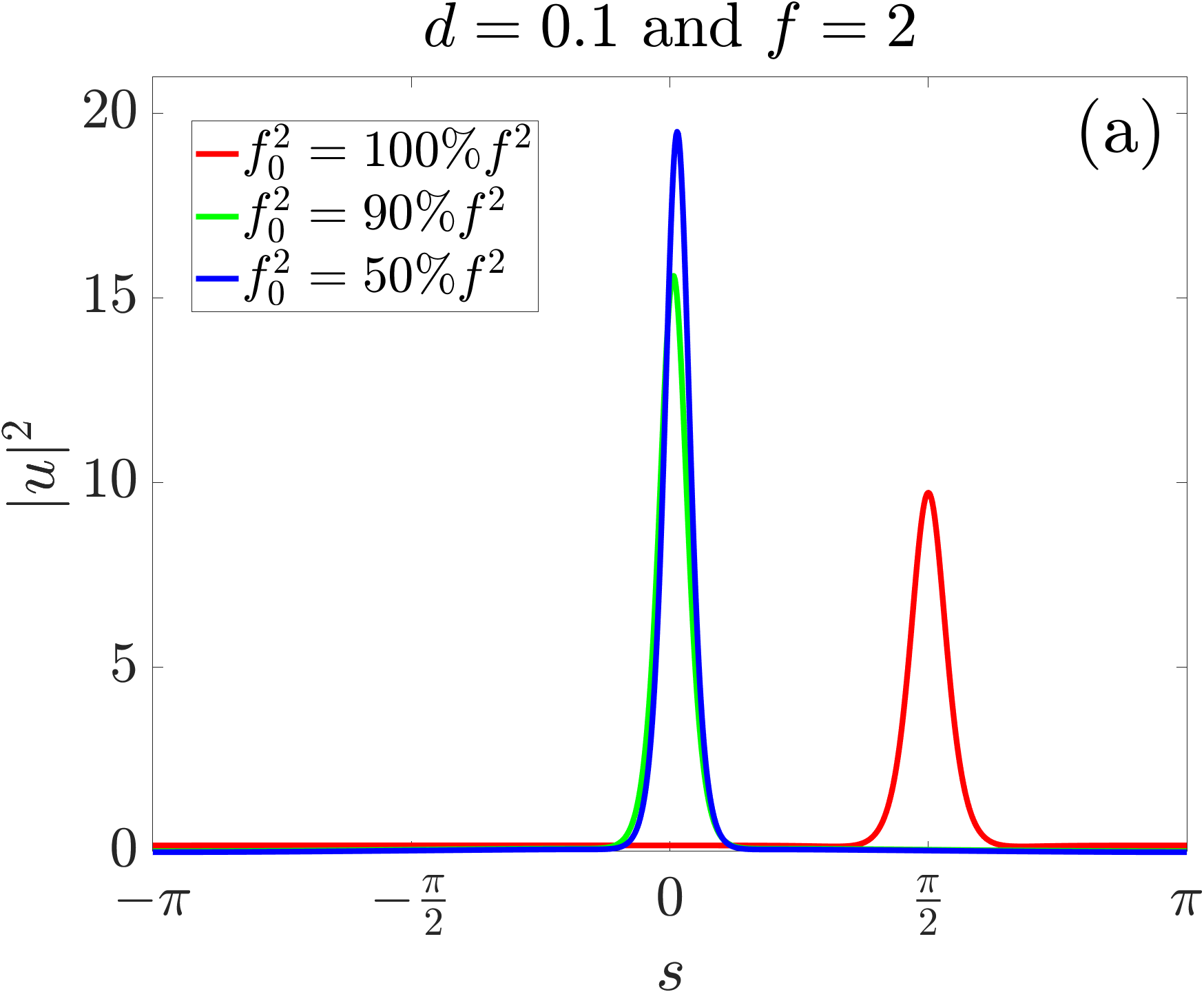} \\[0.5cm]
\includegraphics[width=\columnwidth]{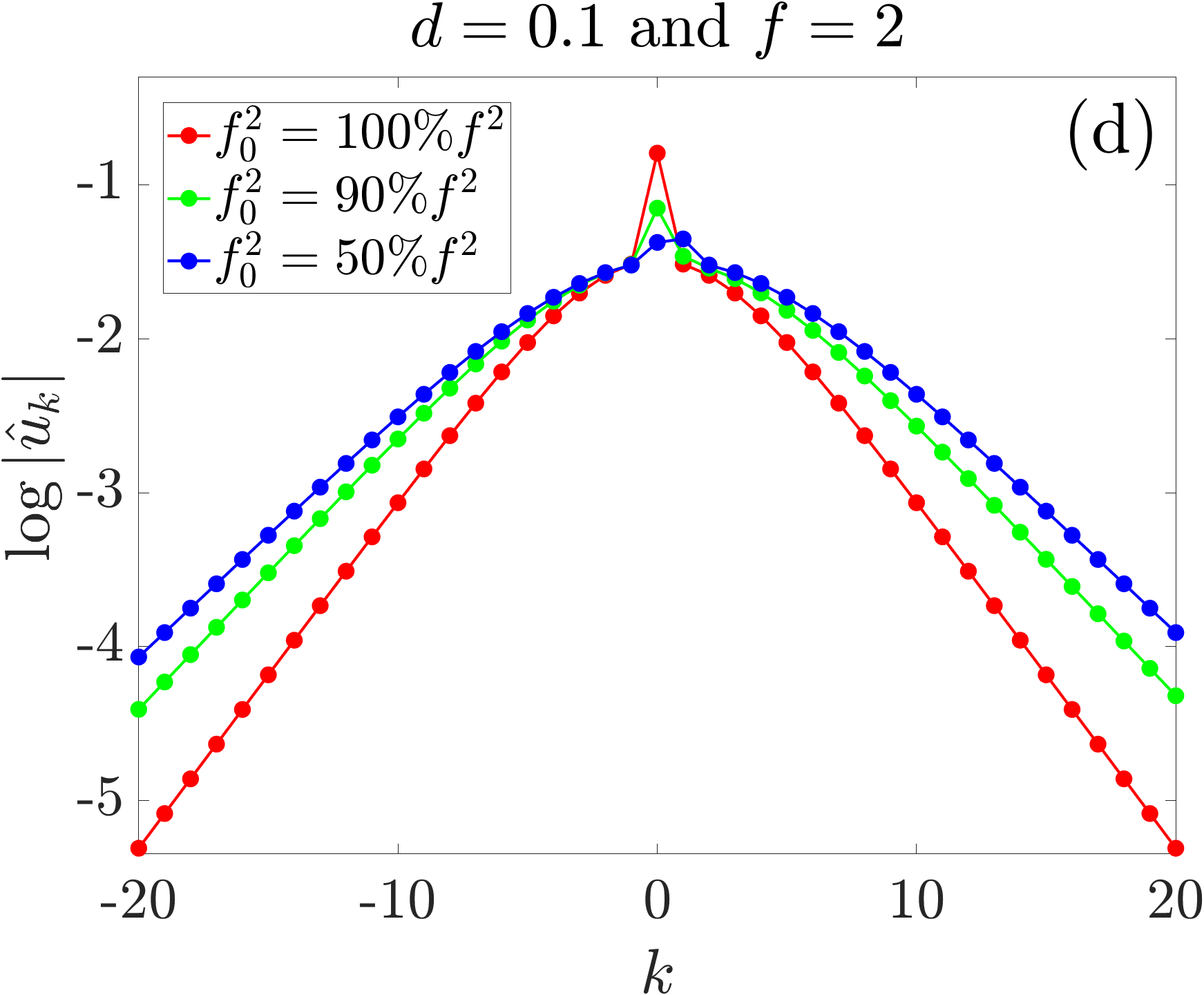}
\end{minipage} \hspace{0.5cm}
\begin{minipage}[t]{0.3\textwidth}
\includegraphics[width=\columnwidth]{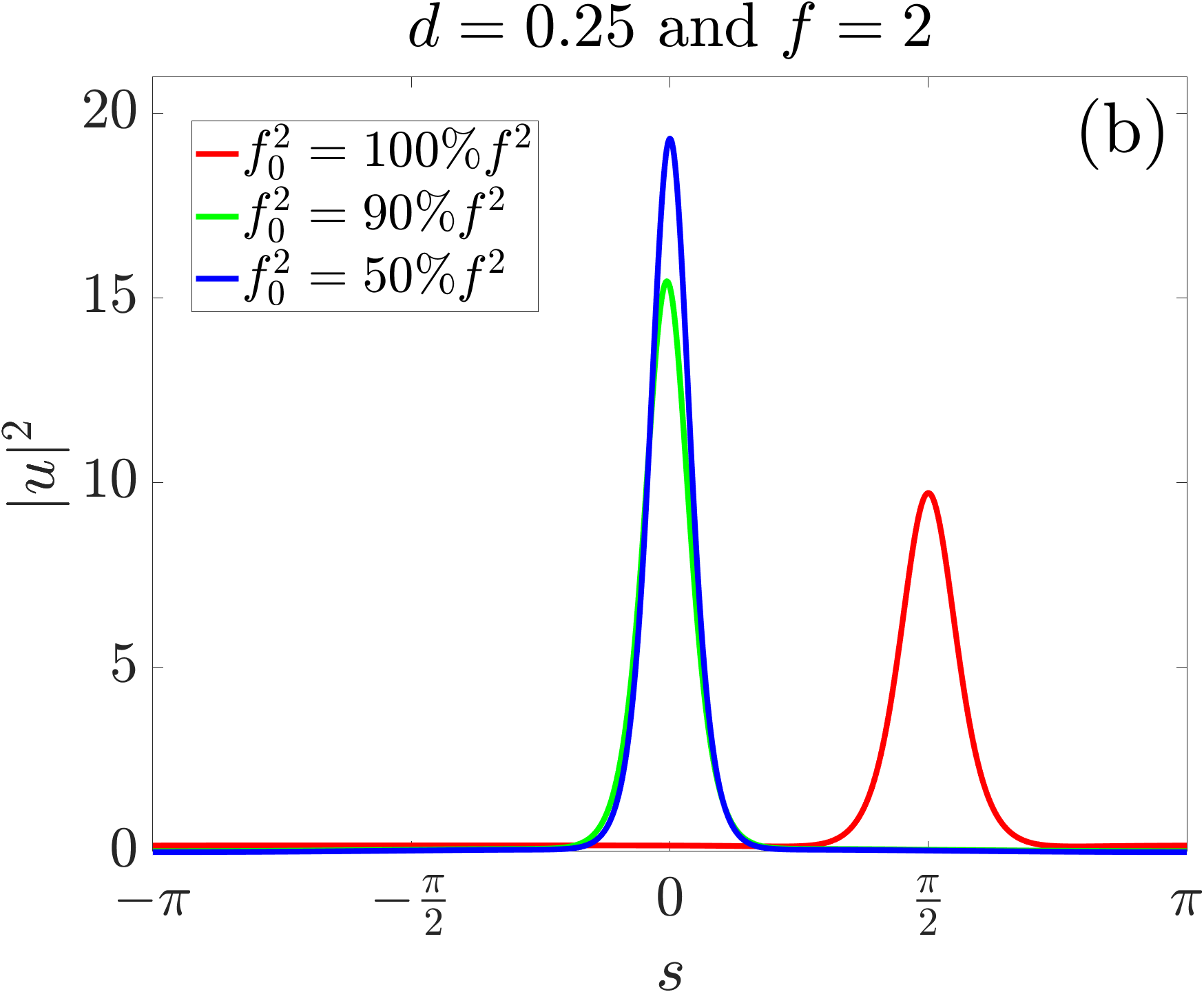} \\[0.5cm]
\includegraphics[width=\columnwidth]{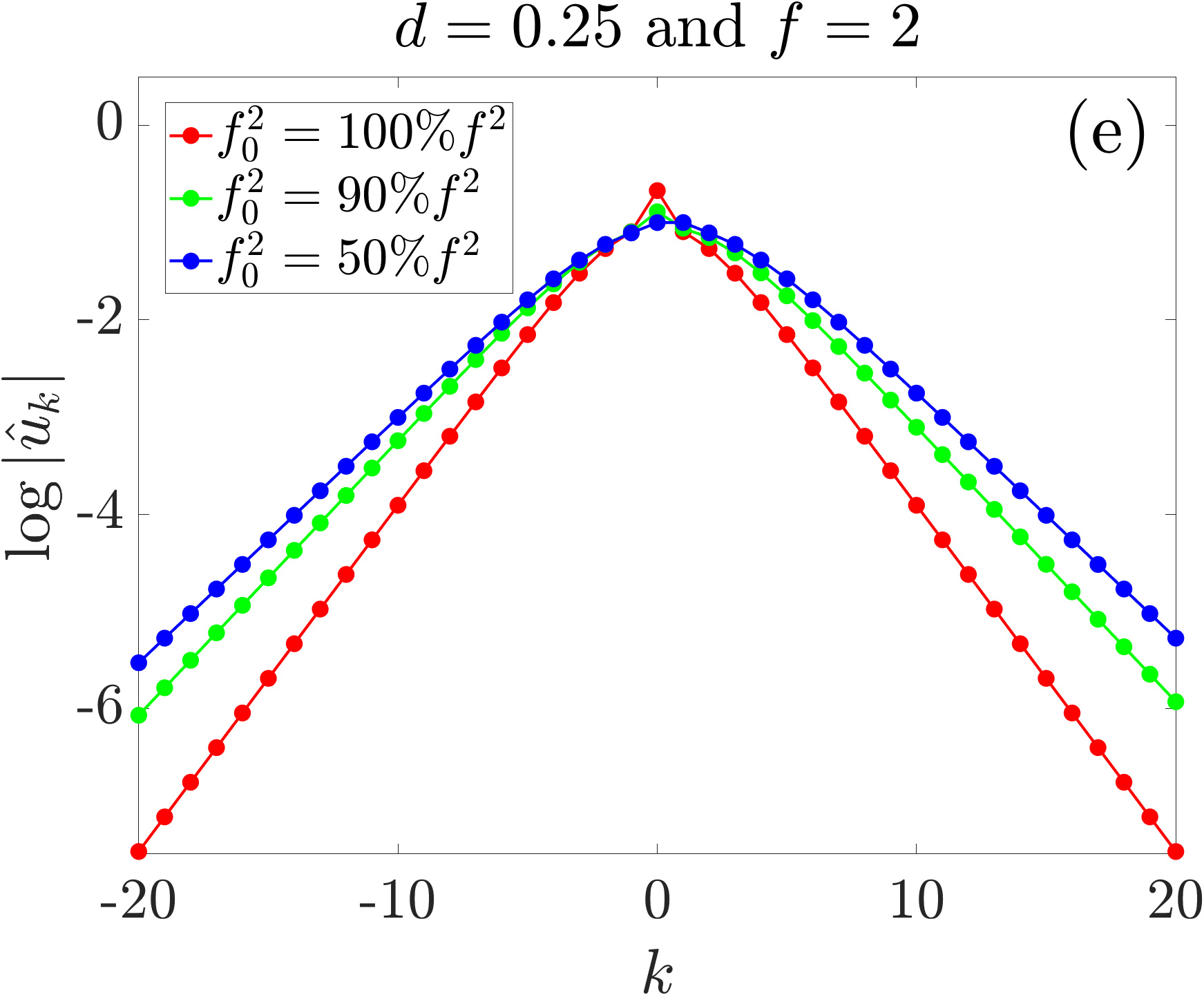}
\end{minipage} \hspace{0.5cm}
\begin{minipage}[t]{0.3\textwidth}
\includegraphics[width=\columnwidth]{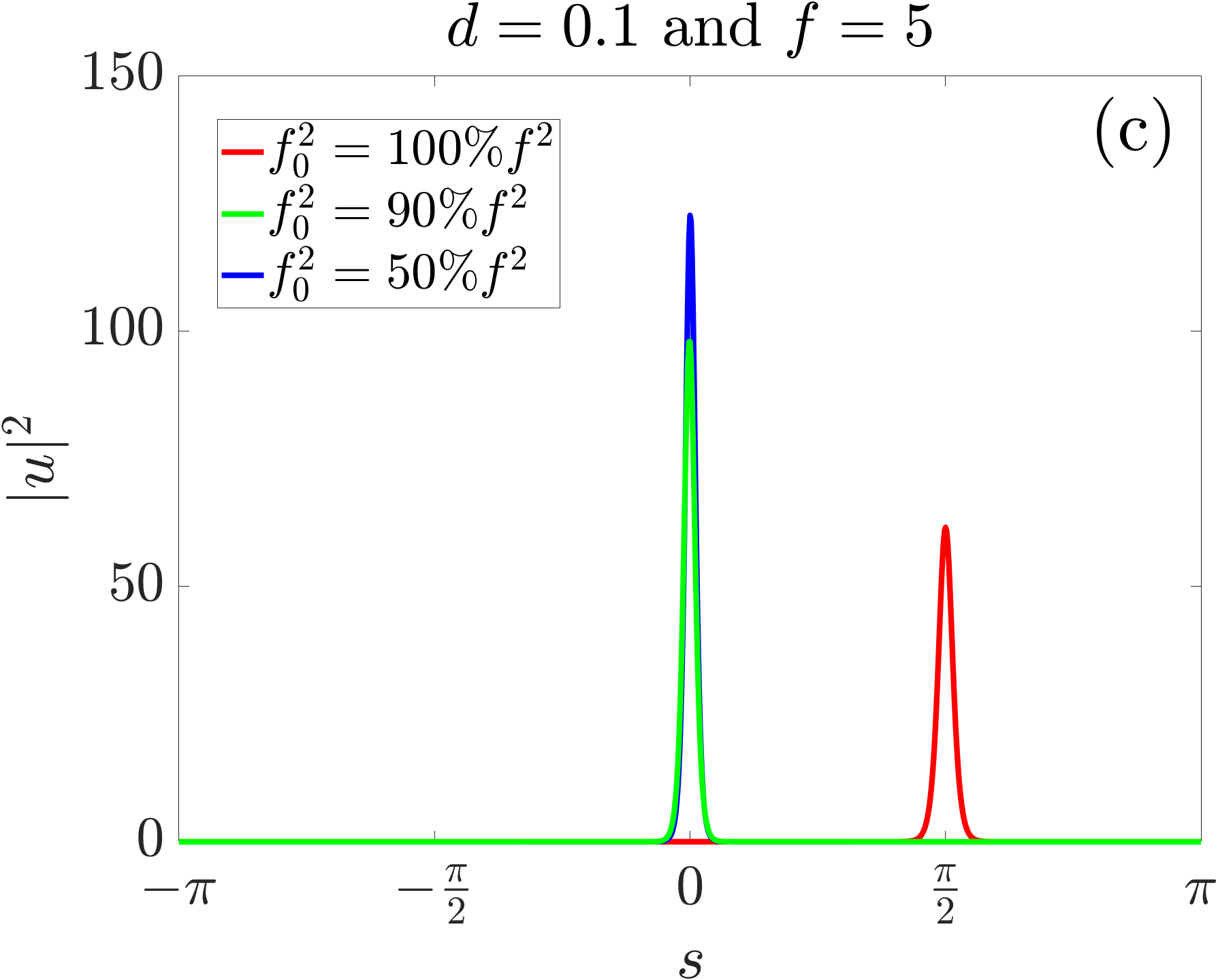} \\[0.5cm]
\includegraphics[width=\columnwidth]{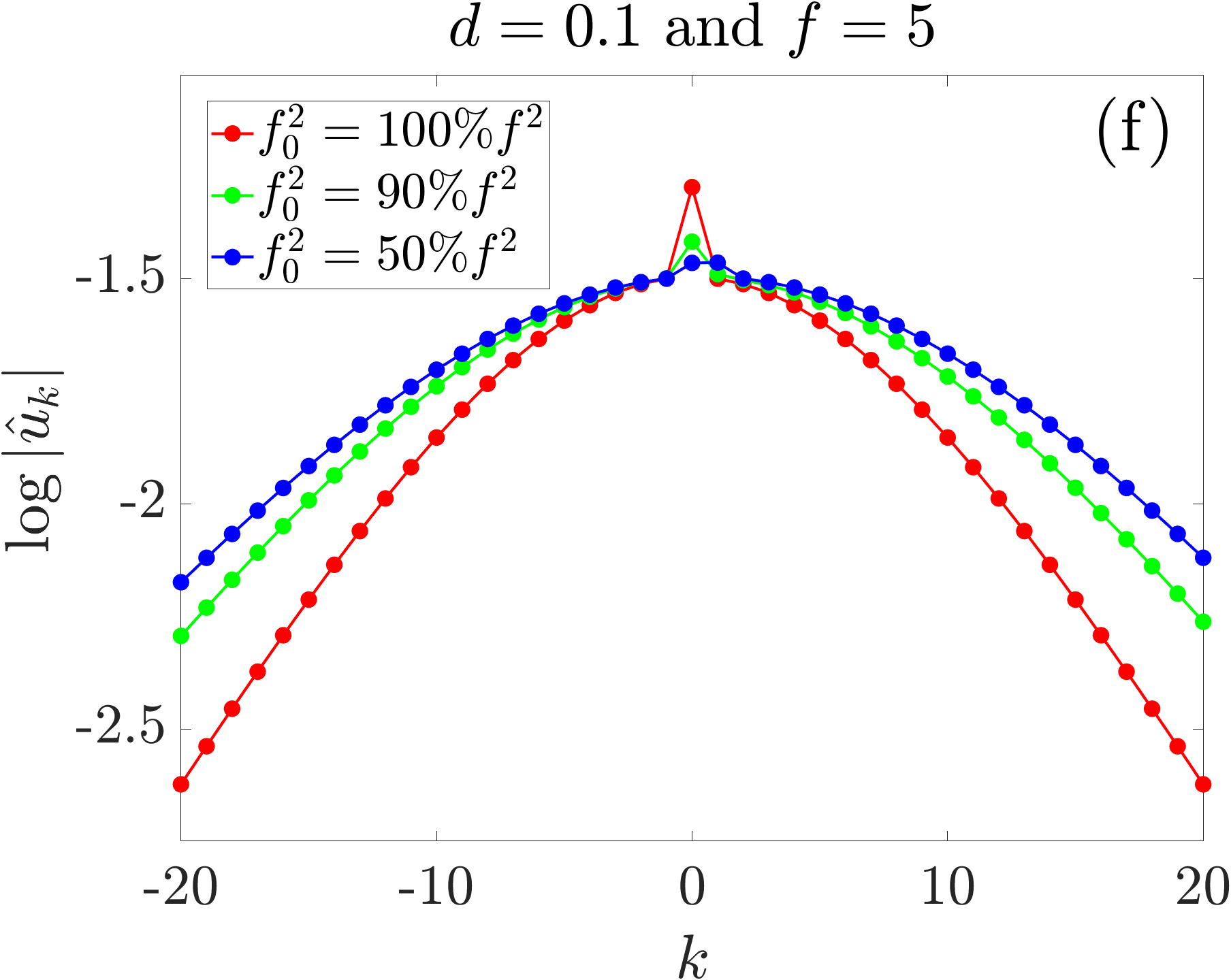}
\end{minipage}
\caption{Spatial and spectral power distributions of optimal solitons for selected values of $\varphi \in [0,\pi/4]$ which correspond to $f_0^2=100\%f^2$, $f_0^2=90\%f^2$ as well as $f_0^2=50\%f^2$ for three different examples. Every column corresponds to one example.}
\label{fig:SolitonsFor100/90/50Per}
\end{figure*}

For each of the three examples from Fig.~\ref{fig:PerformanceMetricsUnderPowerDistribution} and Fig.~\ref{fig:OptimalValuesZetaOmega} we added in Fig.~\ref{fig:SolitonsFor100/90/50Per} plots of the spatial and spectral power distributions of the optimal solitons for selected values of $\varphi\in [0,\pi/4]$. In this range for $\varphi$ we have $f_0, f_1\geq 0$. The particular values of $\varphi$ are chosen as $f_0^2=100\%f^2$ (one mode case), $f_0^2=90\%f^2$ (slight perturbation of the one mode case), and $f_0^2=50\%f^2$ (optimal two mode case). Since for $f_1>0$ there is no shift-invariance in \eqref{TWE} anymore all of the depicted solitons are localized around $s=0$, which is the unique point in the interval $[0,2\pi)$ where the absolute value of the pump term is maximal, i.e., $f_0+f_1=\max_{s\in [0,2\pi)} |\i f_0+\i f_1 \e^{\i s}|$. In other words: the best soliton positions its maximum at the point where the pump has maximal absolute value. 

\medskip

Finally, we explain the symmetry properties of Fig.~\ref{fig:PerformanceMetricsUnderPowerDistribution} and Fig.~\ref{fig:OptimalValuesZetaOmega} from the symmetries of $\eqref{TWE}$. If $u$ solves $\eqref{TWE}$ then $u(\cdot+\pi)$ solves $\eqref{TWE}$ with $f_1$ replaced by $-f_1$ and $-u(\cdot+\pi)$ solves $\eqref{TWE}$ with $f_0$ replaced by $-f_0$. This means that the signs of $f_0$ and $f_1$ are not relevant for the curves in Fig.~\ref{fig:PerformanceMetricsUnderPowerDistribution} and Fig.~\ref{fig:OptimalValuesZetaOmega}. The symmetry with respect to $\pi/4$  of the curves in Fig.~\ref{fig:PerformanceMetricsUnderPowerDistribution} stems from the interchangeability of $f_0$ and $f_1$. Namely, if $u$ solves \eqref{TWE} with given values of $\zeta_0, \omega$ then $v(s) \coloneqq u(-s)\e^{\i s}$ solves 
\begin{align*}
-d v'' +\i \widetilde\omega v'-&(\i-\zeta_1)v-|v|^2 v+\i f_1 +\i f_0 \e^{\i s}=0
\end{align*}
with $\zeta_1=\zeta_0-\omega+d$ and $\widetilde{\omega}=2d-\omega$. Note that the roles of $f_0$ and $f_1$ are now interchanged. The fact that $\zeta_0$ and $\omega$ have changed to $\zeta_1$ and $\widetilde{\omega}$ is not relevant since we optimize anyway in these parameters. Together with (vi) this also explains that the curves in Fig.~\ref{fig:OptimalValuesZetaOmega}(b) and Fig.~\ref{fig:OptimalValuesZetaOmega}(c) are odd with respect to the points $(\pi/4,d)$ and $(\pi/4,0)$, respectively. We also mention that the curves in Fig.~\ref{fig:OptimalValuesZetaOmega}(a) are not symmetric with respect to $\pi/4$ but this is not visible in the plot since the difference $\Delta \zeta=\zeta_0-\zeta_1=\omega-d$ is small compared to $\zeta_0$ and $\zeta_1$.

\section{Trends for varying forcing and varying dispersion}\label{Trends}
\noindent For the results in this section we have carried out a parameter study w.r.t. dispersion $d$ and normalized pump amplitude $f$, considering the behavior of PCE, CBW and FWHM of the best solitons (i.e., minimal FWHM) under optimal power distribution $f_0=f_1=f/\sqrt{2}$. As before, we have fixed the two pumped modes to $k_0=0$ and $k_1=1$. We have considered dispersion parameters $d=0.1,0.15,0.2,0.25$ and normalized total pump amplitude $f \in (0,10]$.  From Section~\ref{OPD} we know that under optimal power distribution the solitons with minimal FWHM arise for $\omega=d$. Using this information we can reduce the optimizations from the heuristic of Section~\ref{Heuristic} to a single optimization step in $\zeta_0$. Since $f_0=f_1$ we see that now PCE is the ratio between
\begin{equation*}
P_{\text{FC}}=\sum_{k\in\Z\setminus\{0,1\}} |\hat u_k|^2 +\frac{1}{2}|\hat u_0|^2+\frac{1}{2}|\hat u_1|^2
\end{equation*}
and the total pump power $f^2$. 

\medskip

\begin{figure*}
\centering
\begin{minipage}[t]{0.3\textwidth}
\includegraphics[width=\columnwidth]{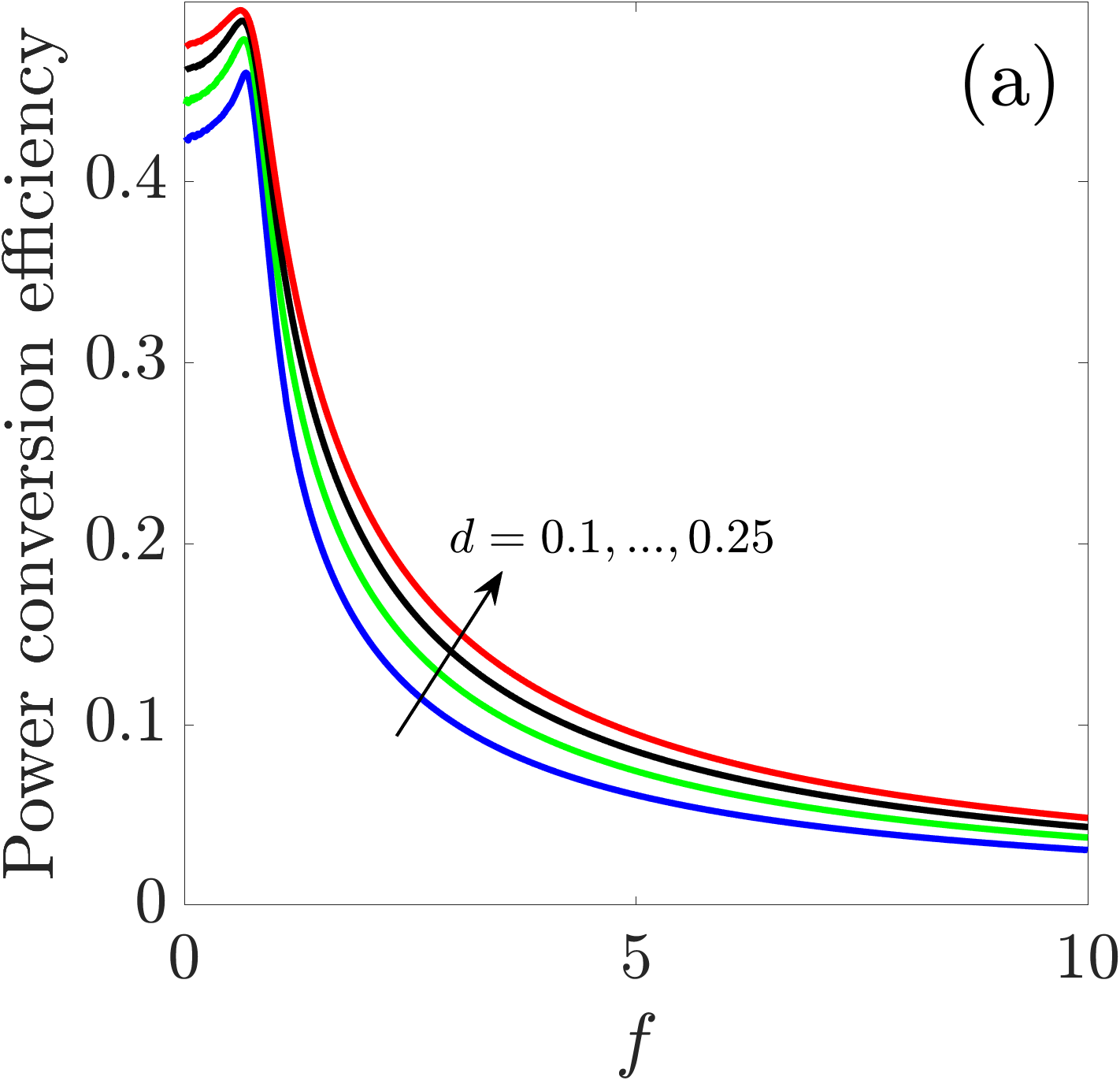}
\end{minipage} \hspace{0.5cm}
\begin{minipage}[t]{0.295\textwidth}
\includegraphics[width=\columnwidth]{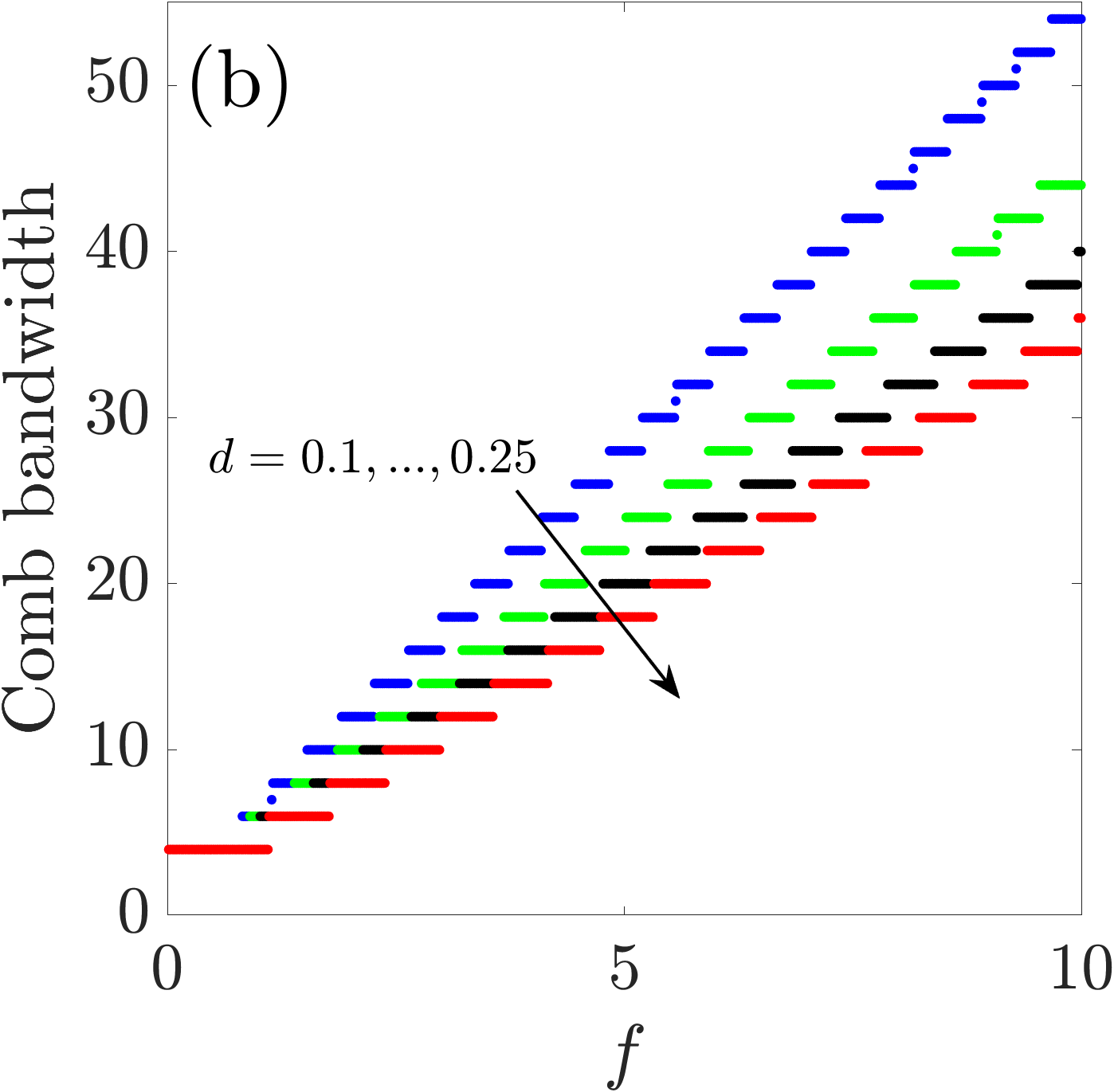}
\end{minipage} \hspace{0.5cm}
\begin{minipage}[t]{0.3\textwidth}
\includegraphics[width=\columnwidth]{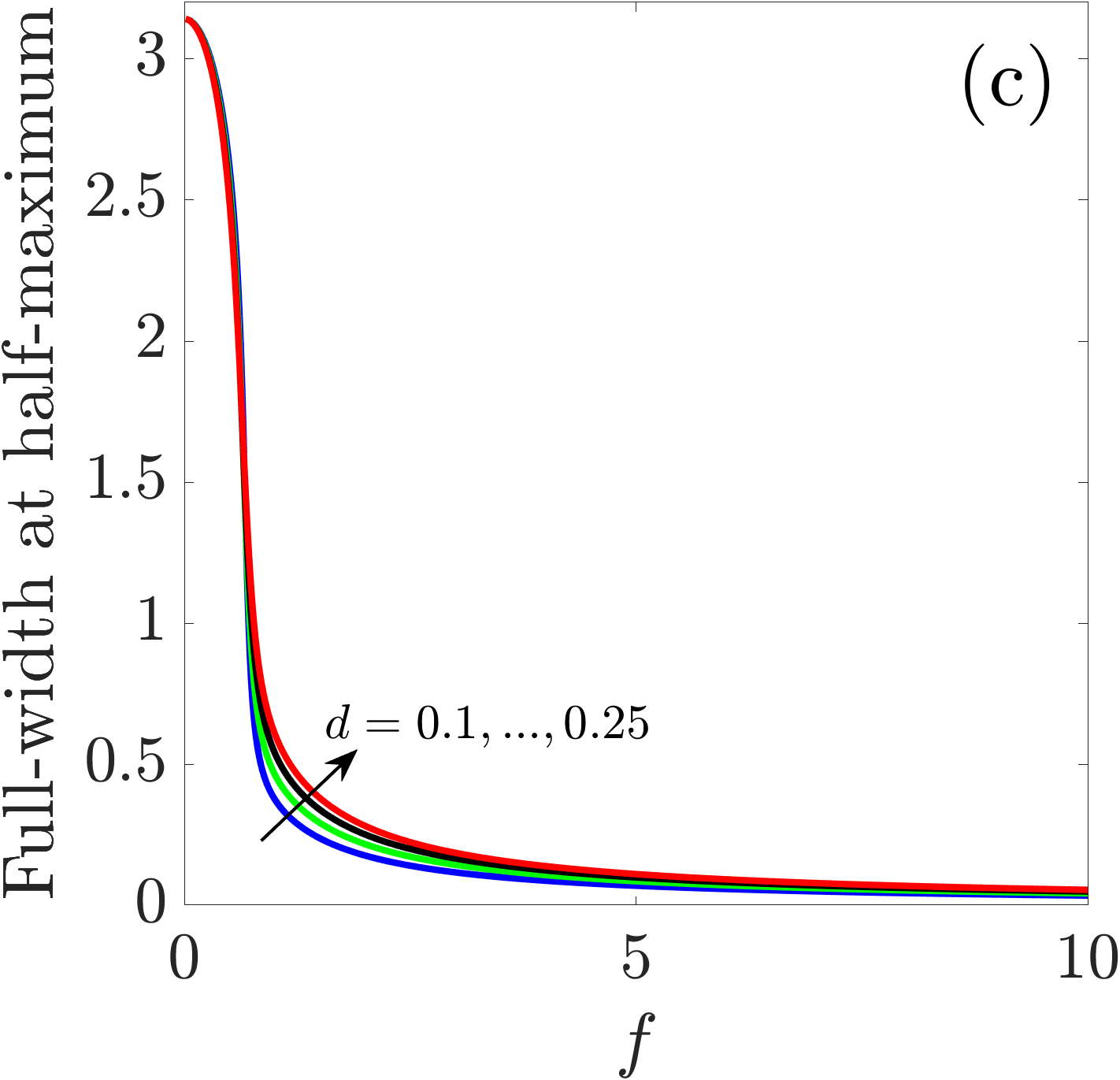}
\end{minipage}
\caption{Power conversion efficiency, comb bandwidth and full-width at half-maximum as a function of the forcing $f$ and dispersion $d=0.1,0.15,0.2,0.25$.}
\label{fig:Trends}
\end{figure*}

The results are shown in Fig.~\ref{fig:Trends}. We observe the following trends: CBW increases whereas FWHM and PCE decrease with increasing forcing $f$. Additionally, one can see that with $d\to 0^+$ once again CBW increases and FWHM, PCE decrease. This observations are in good agreement with the trends from the one mode case, cf. \cite{GaertnerTrochaMandel2018_1000089036}. Further, one can observe in Fig.~\ref{fig:Trends}(c) that FWHM tends to $\pi$ as $f\to 0^+$. This can be understood as follows: as $f\to 0^+$ the solutions of \eqref{TWE} tend to $0$ and behave like the solutions of the linear equation 
$$ -d u'' +\i \omega u'-(\i-\zeta_{0,\text{opt}})u+\i f_0+\i f_1\e^{\i s} = 0.
$$
Since $d=\omega$ for optimal solitons under optimal power distribution $f_0=f_1=f/\sqrt{2}$ the above linear equation is solved by
$$u(s) = \frac{\i f}{\sqrt{2}(\i-\zeta_{0,\text{opt}})}(1+\e^{\i s})$$
and the latter has a FWHM of $\pi$. Similarly, in agreement with Fig.~\ref{fig:Trends}(a), we have
$$\text{PCE}(u) \to \frac{1}{2(1+\zeta_{0,*}^2)} \leq \frac{1}{2}$$
as $f\to 0^+$, where we assume $\zeta_{0,*}=\lim_{f\to 0^+} \zeta_{0,\text{opt}}$. 

\medskip

Finally we mention that the jumps of size two in Fig.~\ref{fig:Trends}(b) could be caused by our choice of the discretization of the $f$-interval $(0,10]$. It is possible that a finer discretization would lead to more plausible jumps of size one. Nevertheless, the finer discretization, which leads to significantly longer run times of the code, has no essential effect on the trends of the curves.

\section{Pumping two arbitrarily distanced modes} \label{arbitrary_k1}
\begin{figure*}
\centering
\begin{minipage}[t]{0.3\textwidth}
\includegraphics[width=\columnwidth]{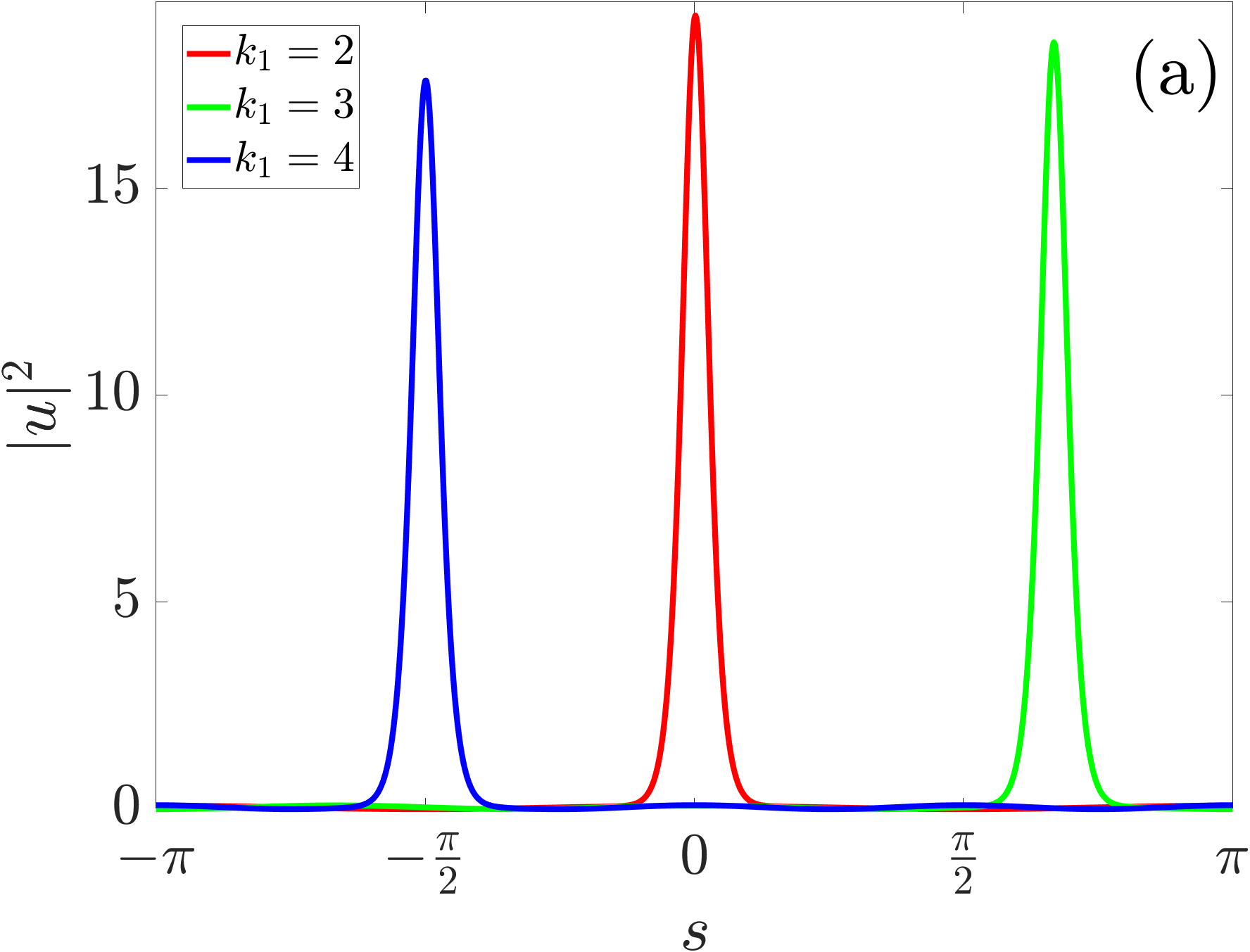}
\end{minipage} \hspace{0.5cm}
\begin{minipage}[t]{0.3\textwidth}
\includegraphics[width=\columnwidth]{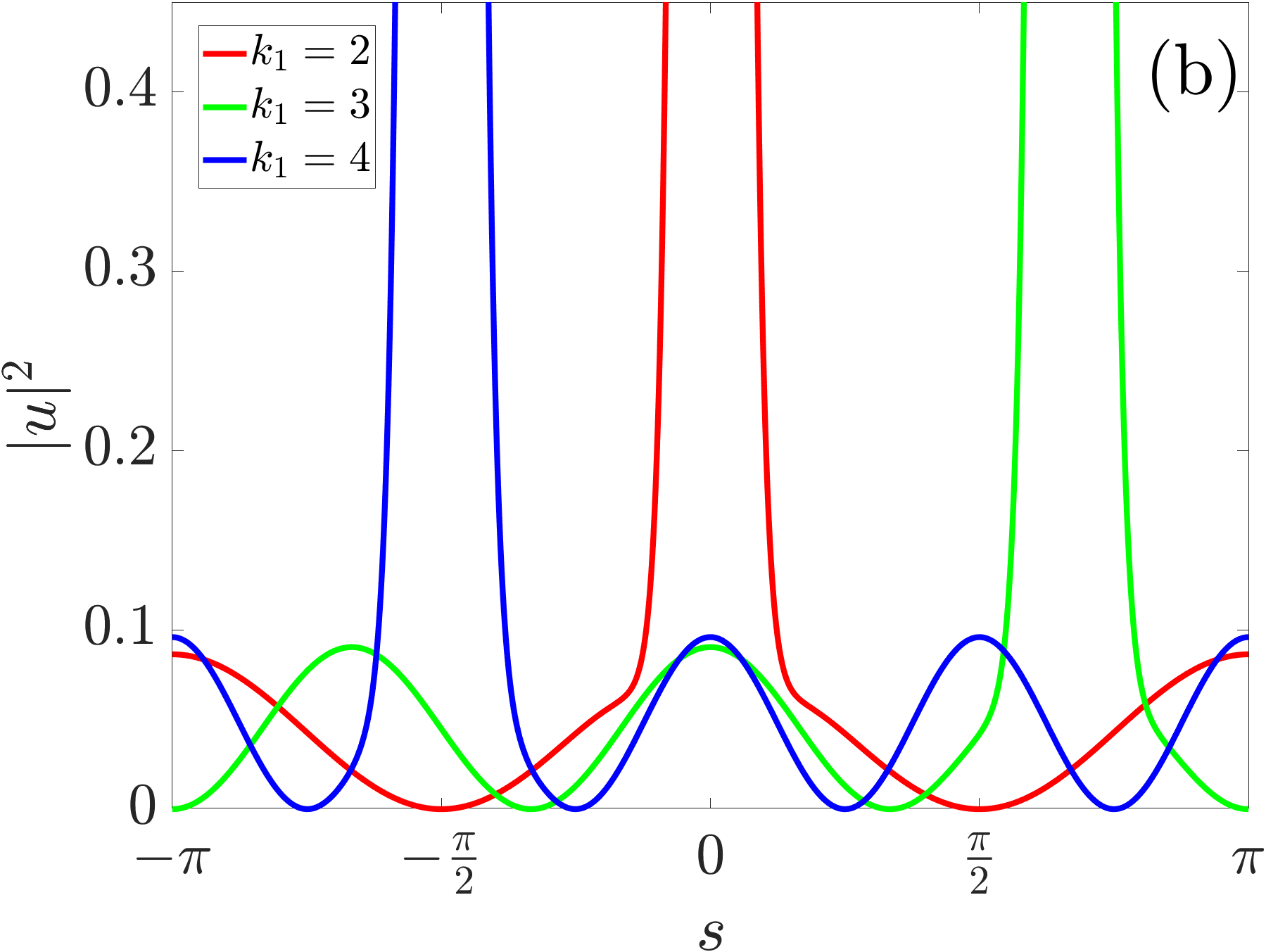}
\end{minipage} \hspace{0.5cm}
\begin{minipage}[t]{0.3\textwidth}
\includegraphics[width=\columnwidth]{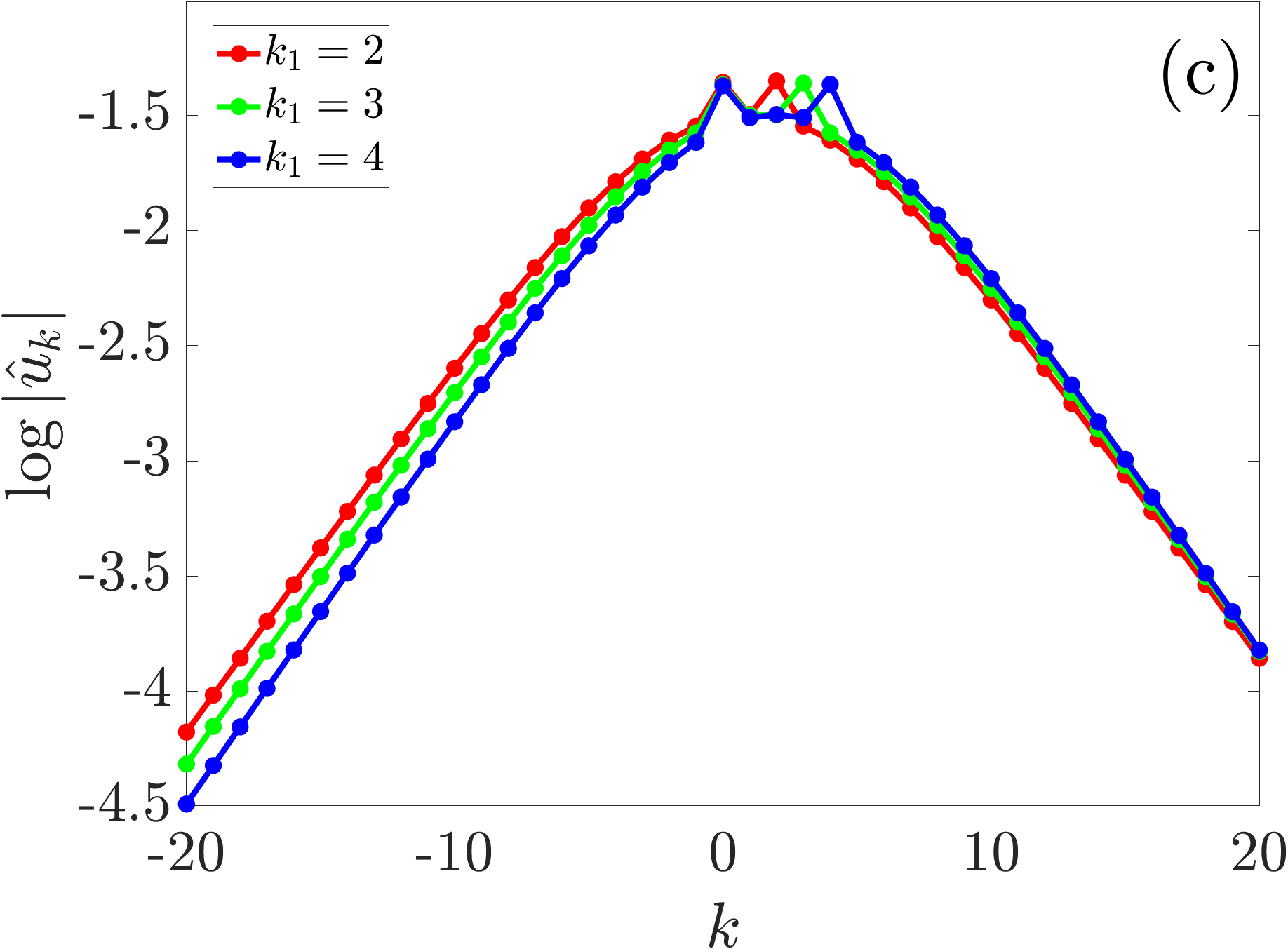}
\end{minipage}
\caption{Spatial and spectral power distributions of the optimal 1-solitons from the case $d=0.1$ and $f_0=f_1=\sqrt{2}$ for $k_1=2,3,4$. Plot (b) is a zoom of plot (a) which highlights the background of the solitons.}
\label{fig:ArbitraryModes}
\end{figure*}

\noindent Also in the case where the pumped modes are $k_0=0$ and $k_1\geq 2$ we have a heuristic algorithm which enables us to identify a 1-soliton with the strongest spatial localization. The algorithm is based on a variant of the one from the case $k_1=1$, cf. Section~\ref{Heuristic}, and details can be found in Appendix C. Applying this algorithm our experiments suggest that the optimal power distribution is again given by the equal distribution $|f_0|=|f_1|$ as in the case $k_1=1$. Moreover, for equal power distribution, $\omega=k_1 d$ turns out to be optimal, which once again translates into equal detuning offsets $\Delta \zeta=0$. In Fig.~\ref{fig:ArbitraryModes}(a) we plotted the spatial power distributions of the optimal 1-solitons from the case $d=0.1$ and $f_0=f_1=\sqrt{2}$ for $k_1=2,3,4$. One can observe that the optimal 1-soliton gets less localized as $k_1$ increases. In Fig.~\ref{fig:ArbitraryModes}(b) we added a zoom-in to better point out the background of the solitons. Since with $u$ also $u(\cdot+2\pi/k_1)$ is a solution of \eqref{TWE} optimal 1-solitons can be shifted by multiples of $2\pi/k_1$. We see that the 1-soliton localizes once again around one of the points where the absolute value of the pump term $\i f_0+\i f_1 \e^{\i k_1 s}$ is maximized. In Fig.~\ref{fig:ArbitraryModes}(c) we added the spectral power distributions of the optimal 1-solitons. Necessarily each comb is peaked at the pumped modes $k_0=0$ and $k_1$.

\section{Summary} \label{summary}
We have considered pumping two different modes for a Kerr nonlinear microresonator with anomalous dispersion. Using numerical path continuation methods we found and tested a heuristic algorithm which allows to find for fixed normalized total pump power the optimal detuning offsets that provide the most localized 1-soliton. The heuristic applies in its simple form to the case of pumping two adjacent modes and in a more refined form (taking bifurcations into account) also to the case of pumping two arbitrarily distanced modes. Optimal 1-solitons appear to be spectrally stable and localize themselves around the intensity maxima of the pump. While it became clear that pumping two modes is always advantageous to pumping one mode, in the case of pumping two adjacent modes we went deeper into the question of how the normalized total input power should be divided into the two pumped modes in order to optimize quality metrics like PCE, CBW, and FWHM. A detailed parameter study shows that the optimal distribution is always the equal distribution $|f_0|=|f_1|=|f|/\sqrt{2}$ with equal detuning offsets. The situation appears to be similar in the case of pumping two arbitrarily distanced modes. Our approach has thus validated the assumptions in \cite{PhysRevA.90.013811}. Finally, we determined trends of PCE, CBW, and FWHM by varying anomalous dispersion and normalized total input power. The trends are in good agreement with the case of pumping only one mode, cf. \cite{GaertnerTrochaMandel2018_1000089036}. Our approach is well-suited to determine and analyze optimal pumping schemes in the case where more than two modes are pumped.

\makeatletter\onecolumngrid@push\makeatother
\makeatletter\onecolumngrid@pop\makeatother

\begin{acknowledgments}
\noindent{Funded by the Deutsche Forschungsgemeinschaft (DFG, German Research Foundation) – Project-ID 258734477 – SFB 1173}
\end{acknowledgments}

\onecolumngrid


\section*{Appendix A: Derivation of the Lugiato-Lefever model for a dual-pumped ring resonator}\label{Derivation of the Lugiato-Lefever model for a dual-pumped ring resonator}
\noindent In this section we derive \eqref{PDE} from a system of coupled mode equations, cf. \cite{Taheri_2017, Herr:12}. When a resonant cavity is pumped by two continuous wave lasers with frequencies $\omega_{p_0}$ and $\omega_{p_1}$ a system of nonlinear coupled mode equations can be used to describe the evolution of the field inside the cavity. The numbering $k$ of the resonant modes in the cavity is relative to the mode $k_0=0$. We use the cold cavity dispersion relation $\omega_k=\omega_0+d_1 k+d_2 k^2$ for the resonant frequencies $\omega_k$, where $d_1$ corresponds to the FSR of the resonator and $2d_2$ to the difference between two neighboring FSRs at the center frequency $\omega_0$.
With $\widetilde k_0,\widetilde k_1 \in \Z$, $\widetilde k_0<\widetilde k_1$, we denote the two pumped modes. If $\hat A_k$ is the mode amplitude of the $k$-th resonant mode normalized such that $|\hat A_k|^2$ is the number of quanta in the $k$-th mode, then the simplified set of equations reads as follows, cf. \cite{Taheri_2017, Herr:12}:
\begin{equation} \label{coupled_mode}
\frac{\partial \hat A_{k}}{\partial t}=-\frac{\kappa}{2}\hat A_k+ \sum_{j=0}^1 \delta_{k \widetilde k_j} \sqrt{\kappa_{\text{ext}}} s_j \e^{-\i(\omega_{p_j}-\omega_{\widetilde k_j})t} \e^{\i \phi_j} 
+\i g \sum\limits_{k'+k''-k'''=k} \hat A_{k'} \hat A_{k''} \bar{\hat{A}}_{k'''} \e^{-\i(\omega_{k'}+\omega_{k''}-\omega_{k'''}-\omega_{k})t}.
\end{equation}
Here, $\kappa=\kappa_0+\kappa_{\text{ext}}$ denotes the cavity decay rate as a sum of intrinsic decay rate $\kappa_0$ and coupling rate to the waveguide $\kappa_{\text{ext}}$, and $\phi_0, \phi_1$ are the initial phases of the pumps. If $P_{\text{in},0}, P_{\text{in},1}$ are the powers of the two input lasers then $s_j=\sqrt{P_{\text{in},j}/\hbar \omega_{\widetilde k_j}}$, $j=0,1$ are the powers coupled to the cavity. The nonlinear coupling coefficient 
$$g=\frac{\hbar \omega_0^2 c n_2}{n_0^2 V_{\text{eff}}}$$ 
denotes a per photon frequency shift of the cavity due to the Kerr nonlinearity and thus describes the strength of the cubic nonlinearity of the system with linear refractive index $n_0$, nonlinear refractive index $n_2$ and effective cavity nonlinear volume $V_{\text{eff}}$. Finally, $c$ is the vacuum speed of light and $\hbar$ the Planck constant. 

\medskip

By using the transformation 
\begin{equation*}
\tilde a(\tau,x) \coloneqq \sqrt{\frac{2g}{\kappa}} \sum_{k\in\Z} \hat A_k\bigg(\frac{2}{\kappa}\tau\bigg) \e^{-\i dk^2 \tau}\e^{\i kx}
\end{equation*}
the system \eqref{coupled_mode} of coupled mode equations may be rewritten in a dimensionless way as a partial differential equation,
\begin{equation} \label{PDE_prelim}
\i \frac{\partial \tilde a}{\partial \tau} =-d \tilde a'' -\i \tilde a-|\tilde a|^2 \tilde a+\i \sum_{j=0}^1 f_j \e^{\i(\widetilde k_j x-\widetilde \nu_j \tau+\phi_j)}, \qquad \tilde a \ 2\pi\text{-periodic in } x,
\end{equation}
where $\tau=\kappa t/2$, $d=2d_2/\kappa$, and $\zeta_j=2(\omega_{\widetilde k_j}-\omega_{p_j})/\kappa$, $\widetilde \nu_j=d \widetilde k_j^2-\zeta_j$, $\eta=\kappa_{\text{ext}}/\kappa$, $f_j=\sqrt{8\eta g/\kappa^2}s_j$ for $j=0,1$. By setting 
$$
a(\tau,x)\coloneqq \e^{-\i(\widetilde k_0 (x+2d \widetilde k_0 \tau-\psi)-\widetilde \nu_0 \tau+\phi_0)} \tilde a(\tau,x+2d \widetilde k_0 \tau-\psi) 
$$
with $\psi=(\phi_1-\phi_0)/k_1$ we find that $a$ satisfies \eqref{PDE} with $k_1=\widetilde k_1-\widetilde k_0$,  $\Delta \zeta=\zeta_0-\zeta_1$ and $\nu_1=\widetilde \nu_1-\widetilde \nu_0 -2d \widetilde k_0 k_1=\Delta\zeta+d k_1^2$. Thus, we can always assume, for simplicity, that the pumped modes are $k_0=0$ and $k_1 \in \N$ and that the initial phase of both pumps is zero. Moreover we see that the change from $\tilde a$ to $a$ shifts the time-dependent Fourier-coefficients from $\hat A_k$ to $\hat A_{k+\tilde k_0}$ and multiplies them with $\e^{-\i(\zeta_0 \tau+\phi_0+k \psi)}$ so that the power in each individual mode is (up to an index shift) preserved.

Finally, let us explain that the intracavity power $P=\sum_{k\in\Z} |\hat u_k|^2=\frac{1}{2\pi}\int_0^{2\pi} \, |u(s)|^2 \, ds$ of a $2\pi$-periodic traveling-wave comb state $u$ cannot exceed the normalized total input power $f^2=f_0^2+f_1^2$. To see this, we multiply the equation \eqref{TWE} for the traveling-wave profile $u$ with $\bar u(s)$ and take the imaginary part to obtain
$$
-d \Im(u''(s) \bar u(s)) + \omega \Re(u'(s) \bar u(s)) - |u(s)|^2 +\Re\bigl((f_0+f_1 \e^{\i k_1 s}) \bar u(s)\bigr) = 0.
$$
Integration over the interval $[0,2\pi]$, using integration by parts for the first term and $\frac{d}{ds} |u(s)|^2 = 2\Re(u'(s)\bar u(s))$ for the second term together with the Cauchy-Schwarz inequality yield
$$
\int_0^{2\pi} |u(s)|^2\,ds = \int_0^{2\pi} \Re\bigl((f_0+f_1 \e^{\i k_1 s}) \bar u(s)\bigr)\,ds \leq \left(\int_0^{2\pi} |u(s)|^2\,ds\right)^{1/2} \sqrt{2\pi}(f_0^2+f_1^2)^{1/2}
$$
and hence $\frac{1}{2\pi}\int_0^{2\pi} |u(s)|^2\,ds \leq f_0^2+f_1^2$.

\section*{Appendix B: Detailed explanation of the heuristic for finding localized solitons in the case of pumping two adjacent modes} \label{AppendixDetailedExplanation}
\noindent Here we explain in detail the heuristic algorithm mentioned in Section~\ref{Heuristic} for finding strongly localized solutions of \eqref{TWE} in the case of anomalous dispersion $d>0$, where two adjacent modes are pumped, i.e. the pumped modes are $k_0 = 0$ and $k_1 = 1$. We recall that the parameters $d>0$, $k_1=1$, $f_0$ and $f_1$ are fixed and that the goal is to find optimally localized solutions by varying the parameters $\zeta_0$ and $\omega$ since they can be influenced by the choice of the pump frequencies $\omega_{p_0}$ and $\omega_{p_1}$ through the relation 
$$
\zeta_0 = \frac{2}{\kappa}\bigl(\omega_0-\omega_{p_0}\bigr), \qquad \omega = \frac{2}{\kappa}\bigl(\omega_0-\omega_{p_0}-(\omega_1-\omega_{p_1})+d_2\bigr).
$$
Without loss of generality we assume $0<f_1 \leq f_0$. The heuristic algorithm consists of the following steps. For all our computations we carried it out by using \texttt{pde2path}. 

\medskip

\emph{
\begin{itemize}
\item[] Step 0: Initialize with $f_1^{(0)}=0$, $\zeta_0^{(0)}=2+f^2$, $\omega^{(0)}=0$, find $u_0=$ constant solution of \eqref{TWE} and set $j=1$
\item[] Step 1 ($f_1$-continuation): with $\zeta_0^{(0)}$, $\omega^{(j-1)}$ start from $f_1^{(0)}$ and continue $u_0$ in $f_1$-parameter until desired value $f_1$ is reached, keep solution $T_{j}$
\item[] Step 2 ($\zeta_0$-optimization): with $\omega^{(j-1)}$, $f_1$ start from $\zeta_0^{(0)}$ and continue $T_{j}$ in $\zeta_0$-parameter until $1$-solitons have been exhausted, find optimal $\zeta_0^{(j)}$, keep optimal soliton $A_j$
\item[] Step 3 ($\omega$-optimization): with $\zeta_0^{(j)}$, $f_1$ start from $\omega^{(j-1)}$ and continue $A_j$ in $\omega$-parameter on closed loop, find optimal $\omega^{(j)}$, keep optimal soliton $B_j$
\item[] $j\to j+1$, return to Step 1 unless desired accuracy achieved
\end{itemize}
}

\medskip

Now we comment on the individual steps. 

\medskip 

\noindent
\emph{Step 0:} The algorithm starts by choosing suitable initial values for the parameters $f_1$, $\zeta_0$ and $\omega$. For the values of $f_1^{(0)}=0$ and $\zeta_0^{(0)}=2+f^2$ we can determine a constant solution $u_0$ of \eqref{TWE}. It satisfies 
\begin{equation*}
0=-(\i-\zeta_0^{(0)})u_0-|u_0|^2 u_0+\i f_0.
\end{equation*}
If we choose $\zeta_0^{(0)}$ sufficiently large (in all numerical experiments $\zeta_0^{(0)}=2+f^2$ was sufficient) then $u_0$ is uniquely determined. Since the dispersion $d$ and the difference of the normalized offsets between the pump frequencies $\omega_{p_i}$ and the resonant frequencies $\omega_i$, $i=0,1$ turn out to be rather small we expect that also $\omega=\Delta\zeta+d$ is rather small. Therefore the initial value $\omega^{(0)}=0$ is feasible. 

\medskip

\noindent
\emph{Step 1 ($f_1$-continuation):}
Starting from $\zeta_0^{(0)}$, $\omega^{(j-1)}$ and $f_1^{(0)}$ \texttt{pde2path} performs a continuation algorithm in the $f_1$-parameter. With the side constraint of always solving \eqref{TWE} the trivial state $u_0$ is continued numerically w.r.t. the $f_1$-parameter until the desired value $f_1$ is reached for the first time. Although the starting point $u_0$ is independent of $\omega$ the continuation w.r.t. the $f_1$-parameter is sensitive to the current value of $\omega$. 

\medskip

\noindent
\emph{Step 2 ($\zeta_0$-optimization):} Now that the $f_1$-parameter has reached its correct value we freeze the values of $\omega^{(j-1)}$ and $f_1$ and start the optimization w.r.t. the $\zeta_0$-parameter from $\zeta_0^{(0)}=2+f^2$.  Starting from $\zeta_0^{(0)}$, the continuation of solutions of \eqref{TWE} w.r.t. $\zeta$ first provides almost trivial solutions until they develop into $1$-solitons followed by less localized higher solitons. From the point of view of FWHM-minimization it is therefore reasonable to continue from $\zeta_0^{(0)}$ until the part of the branch containing $1$-solitons has been exhausted. Along this part of the branch the optimal solution $A_j$ of \eqref{TWE} with the minimal FWHM together with the optimal parameter value $\zeta_0^{(j)}$ are kept. 

\medskip

\noindent
\emph{Step 3 ($\omega$-optimization):} Now we freeze $f_1$ and $\zeta_0^{(j)}$. The optimal point $A_j$ from the previous step serves as starting point for the subsequent $\omega$-continuation. Beginning with $\omega^{(j-1)}$ the continuation of solutions to \eqref{TWE} in the $\omega$-parameter always delivers a closed loop. From this closed $\omega$-loop the optimal solution $B_j$ of \eqref{TWE} with the minimal FWHM together with the optimal parameter value $\omega^{(j)}$ is kept.

\medskip

At this point the algorithm is not yet finished since a single optimization in $\zeta_0$ followed by a single optimization in $\omega$ is not an adequate substitute for a continuous two-parameter optimization in $\zeta_0$ and $\omega$. Therefore, the algorithm has to be suitably iterated until a desired accuracy (measured in the deviations of $A_j, B_j$ from its predecessors $A_{j-1}, B_{j-1}$) is achieved. One might think of using $B_j$ as starting point for the next $\zeta_0$-continuation. However, this turns out to be non-optimal in some cases because after the update of $\omega^{(j)}$ the solution $B_j$ no longer lies on a $\zeta_0$-branch that leads to an optimal FWHM. Instead, our strategy is to only keep the value $\omega^{(j)}$, forget the solution $B_j$ and iterate by starting again with Step~1 instead of Step~2, i.e., by starting the $f_1$-continuation from $f_1^{(0)}=0$ (with the by now updated value of $\omega$). The subsequent $\zeta_0$-continuation of Step~2 provides a $\zeta_0$-branch with apparently smaller FWHM.

\medskip

\begin{figure*}
\centering
\begin{minipage}[t]{0.3\textwidth}
\includegraphics[width=\columnwidth]{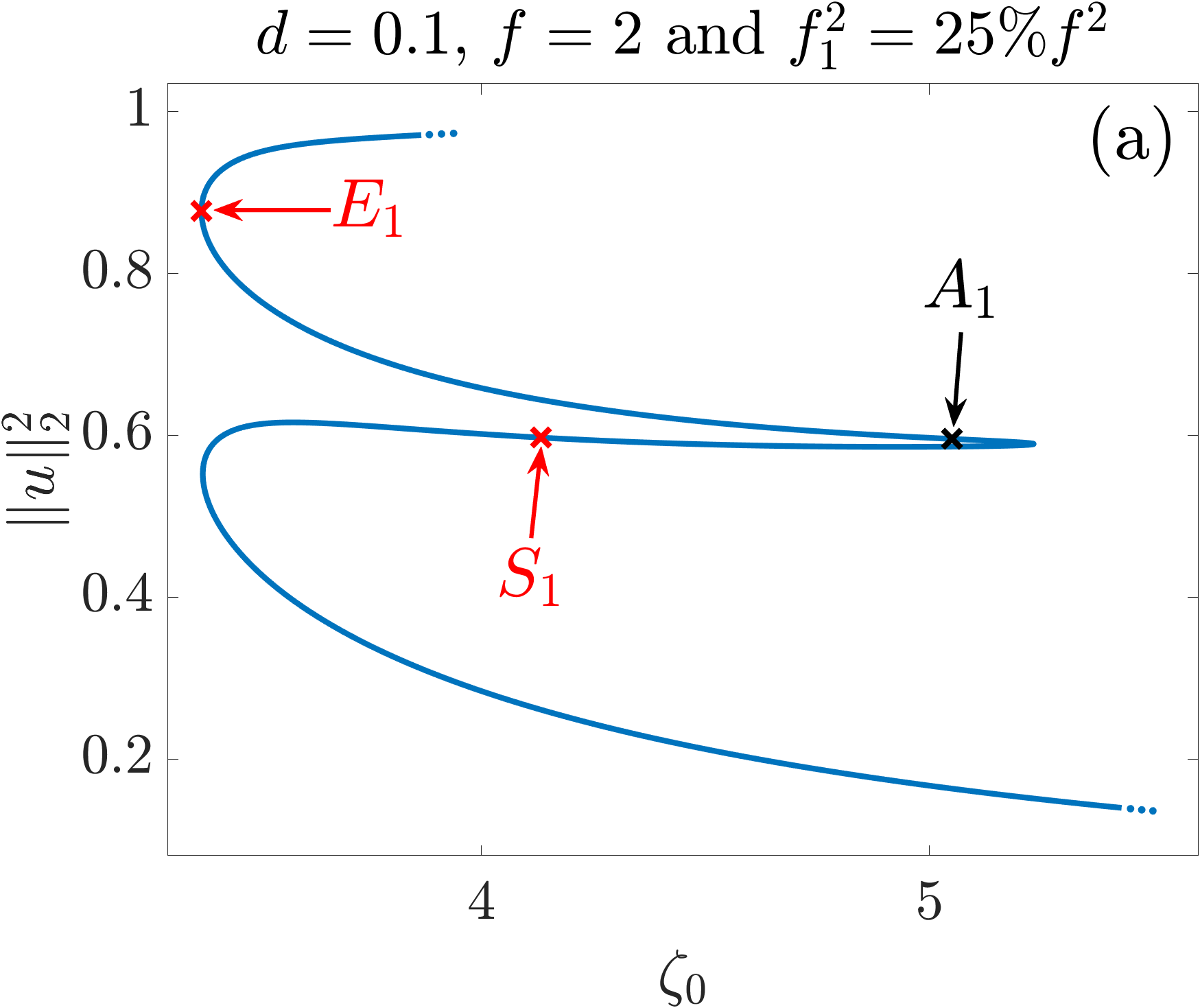} \\[0.5cm]
\includegraphics[width=\columnwidth]{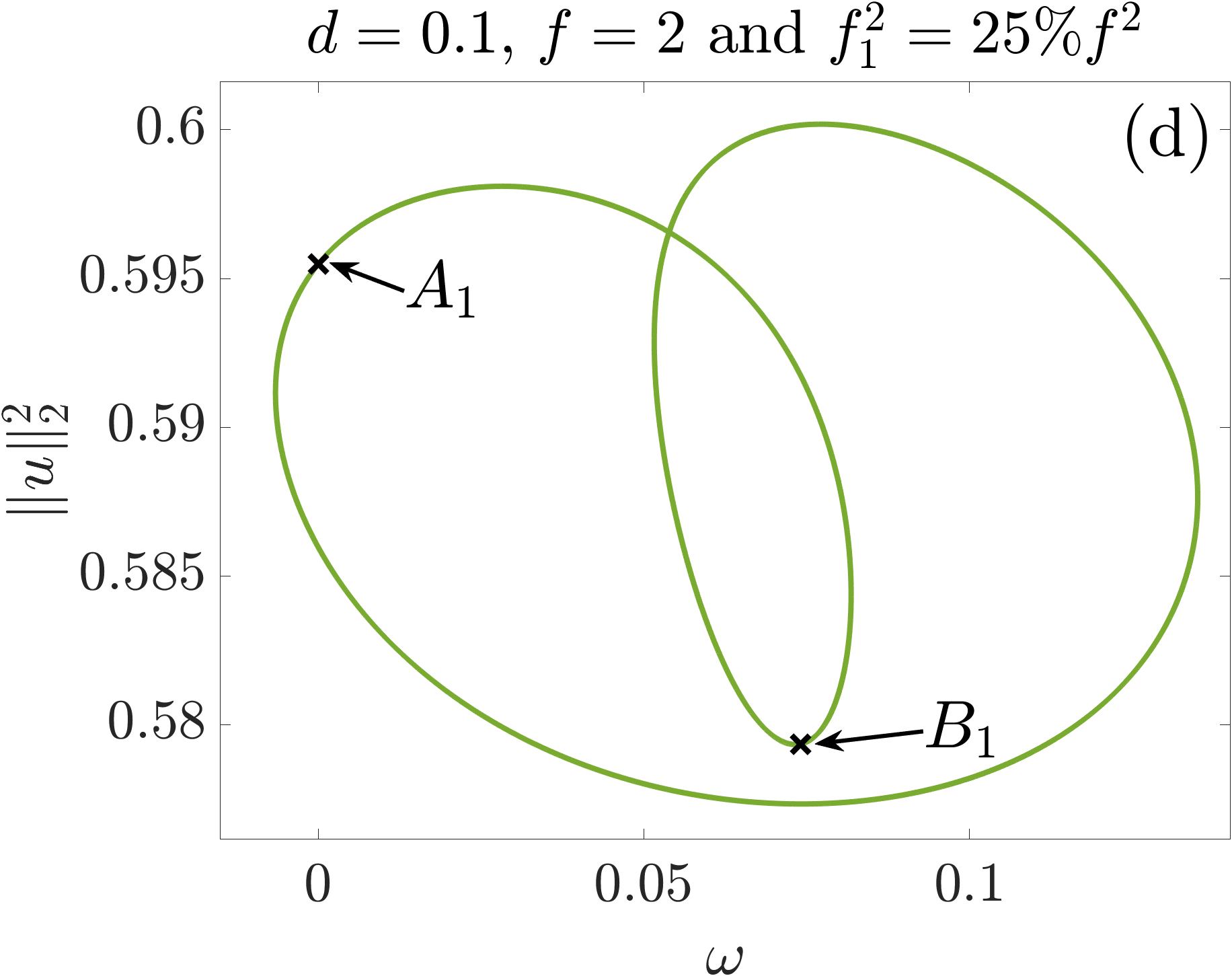} \\[0.5cm]
\includegraphics[width=\columnwidth]{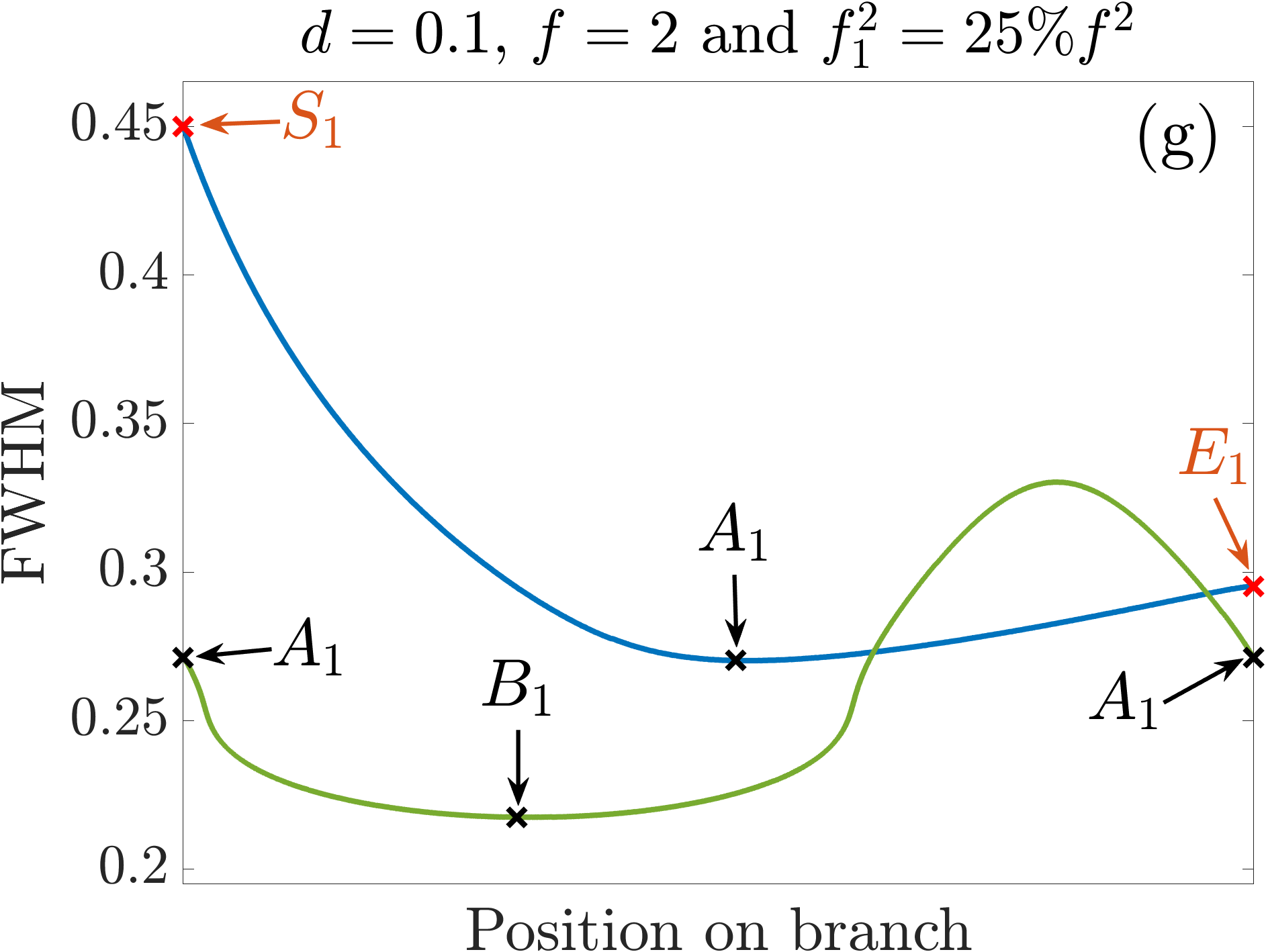}
\end{minipage} \hspace{0.5cm}
\begin{minipage}[t]{0.3\textwidth}
\includegraphics[width=\columnwidth]{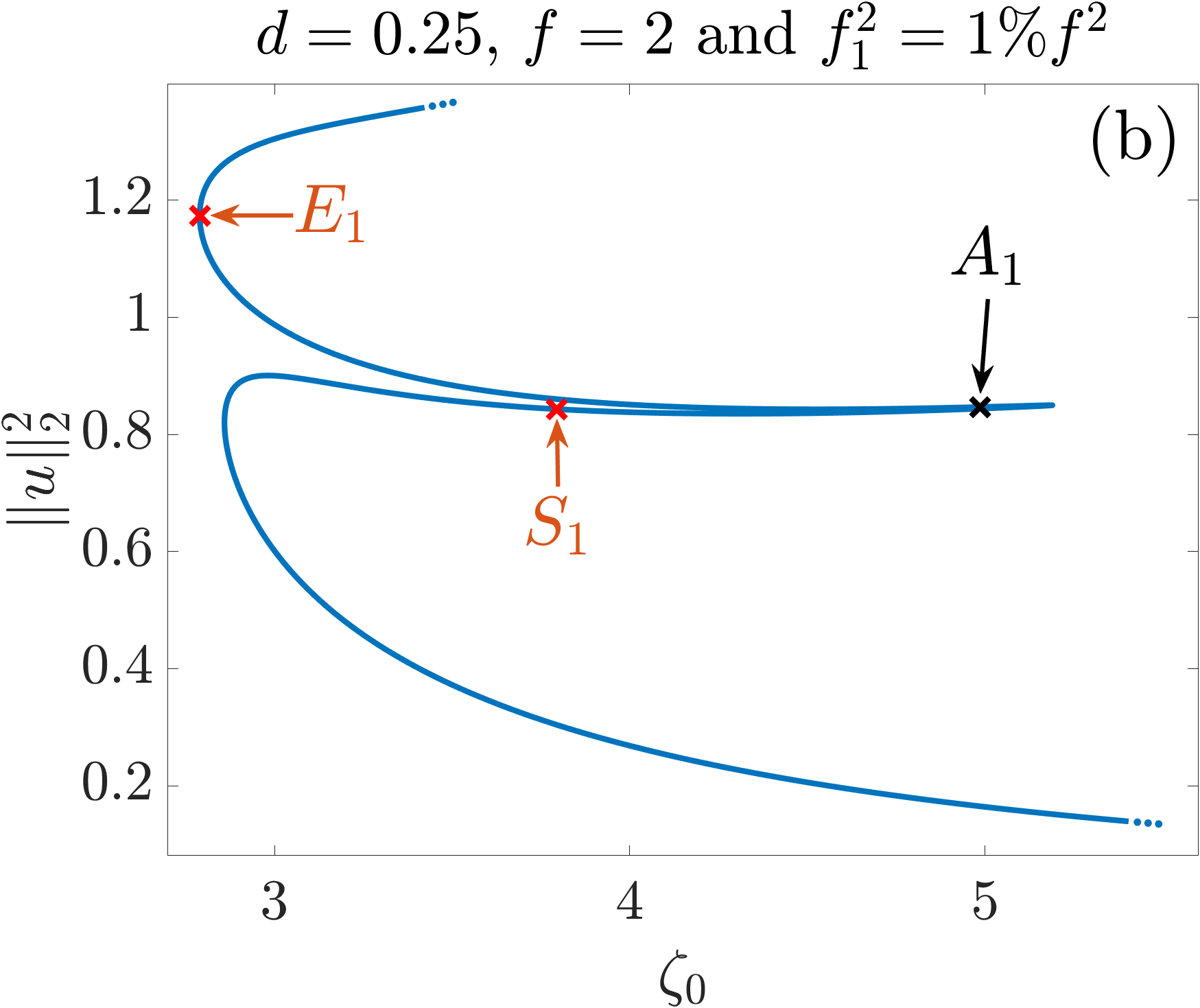} \\[0.5cm]
\includegraphics[width=\columnwidth]{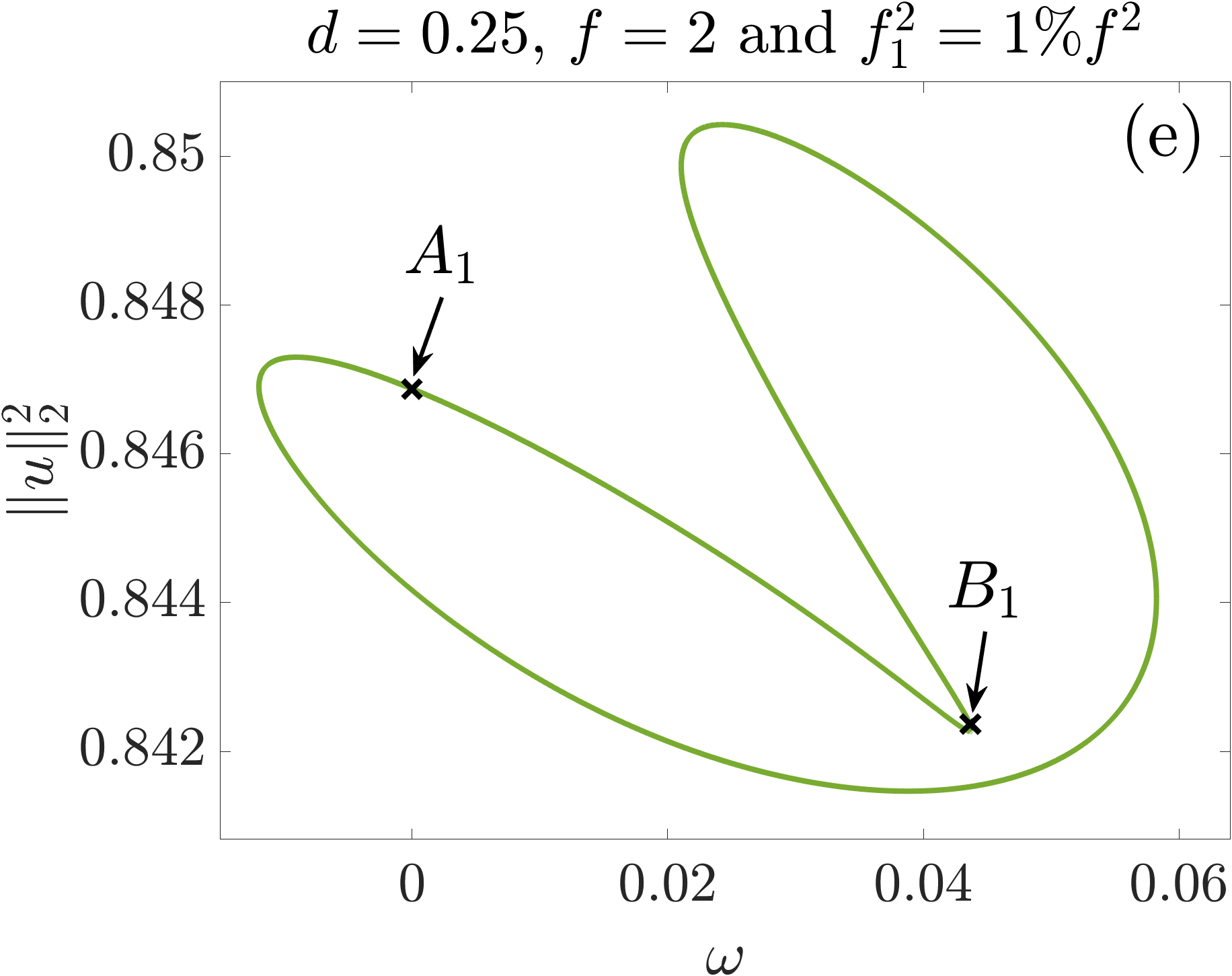} \\[0.5cm]
\includegraphics[width=\columnwidth]{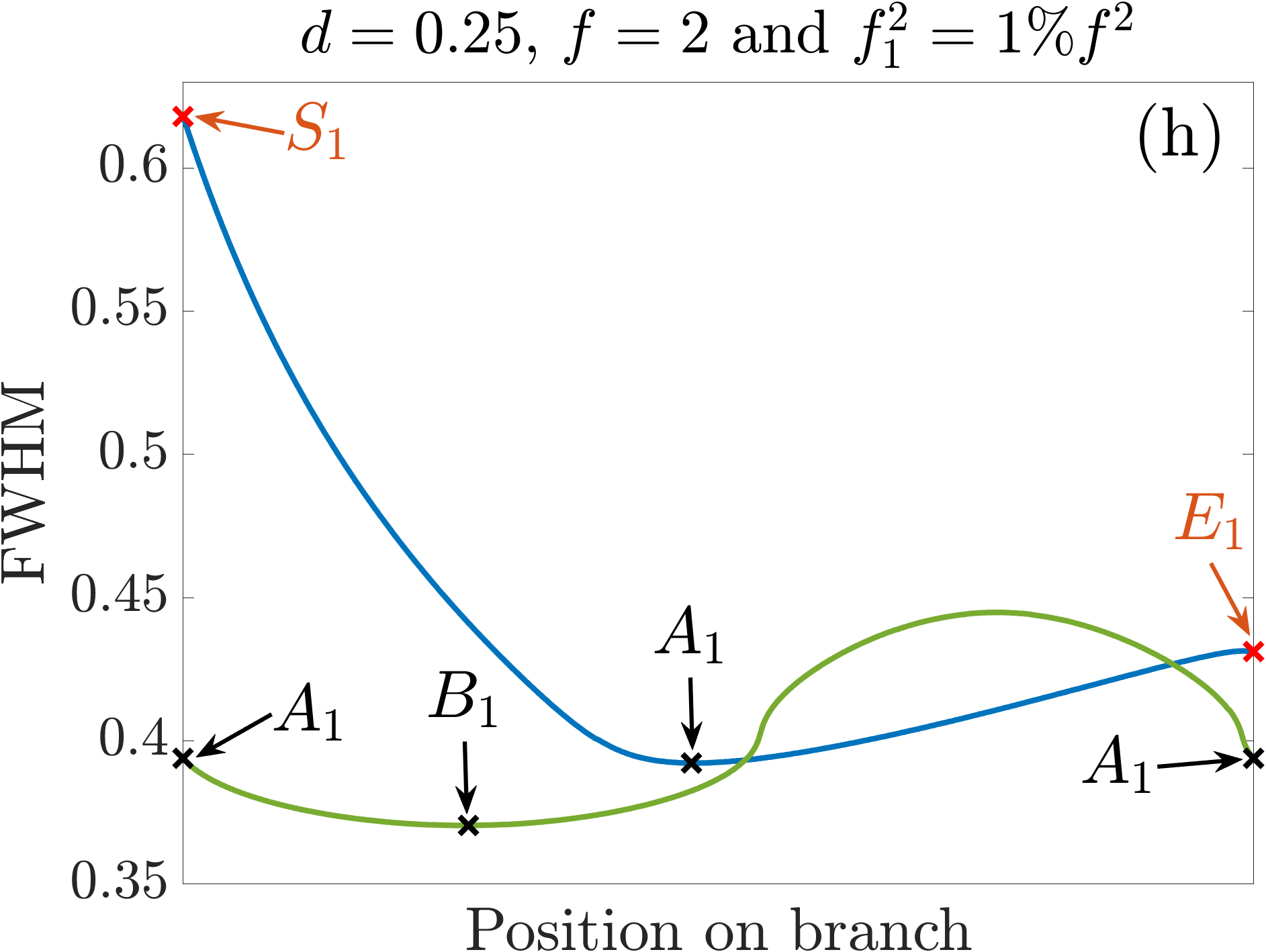}
\end{minipage} \hspace{0.5cm}
\begin{minipage}[t]{0.3\textwidth}
\includegraphics[width=\columnwidth]{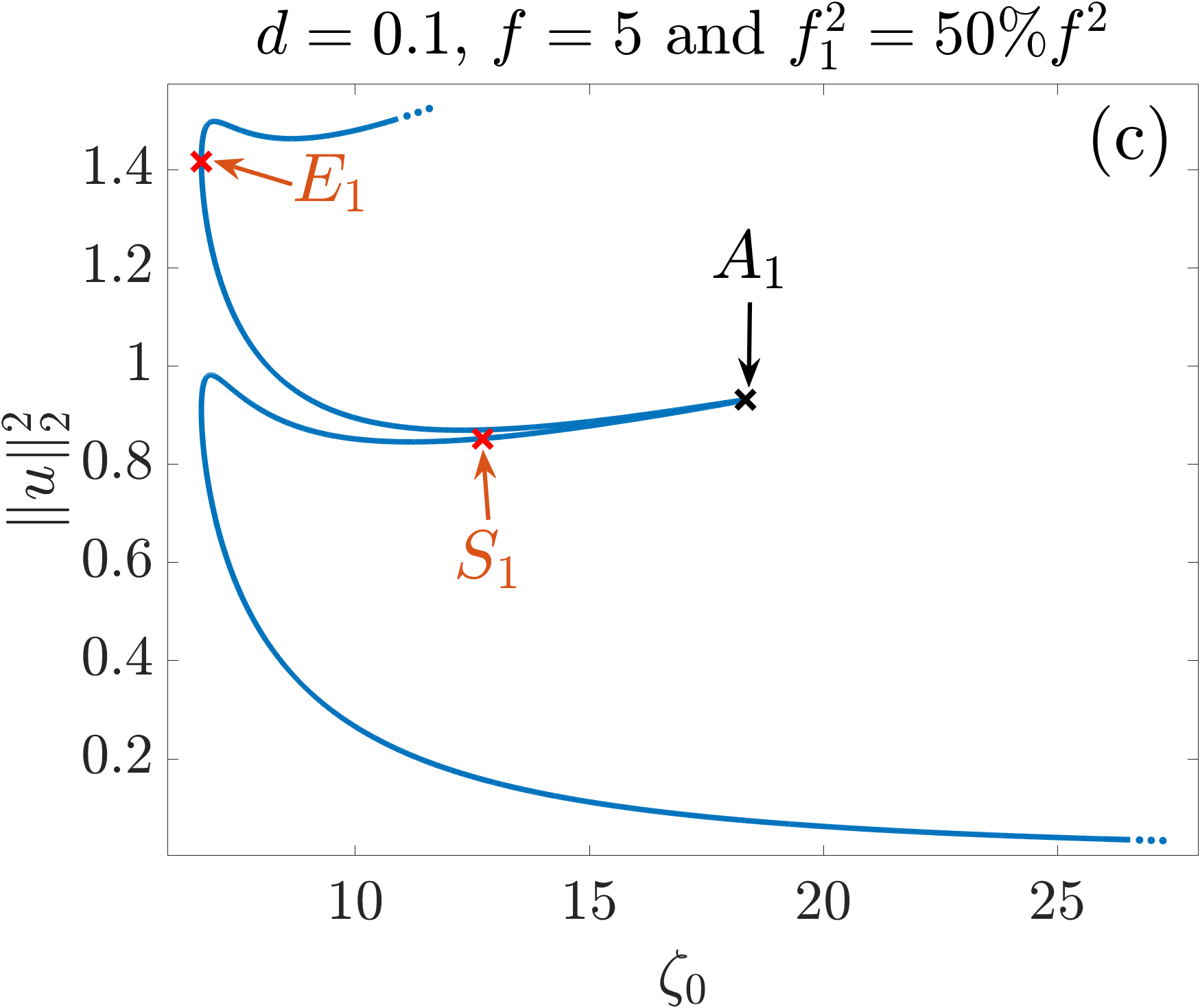} \\[0.5cm]
\includegraphics[width=\columnwidth]{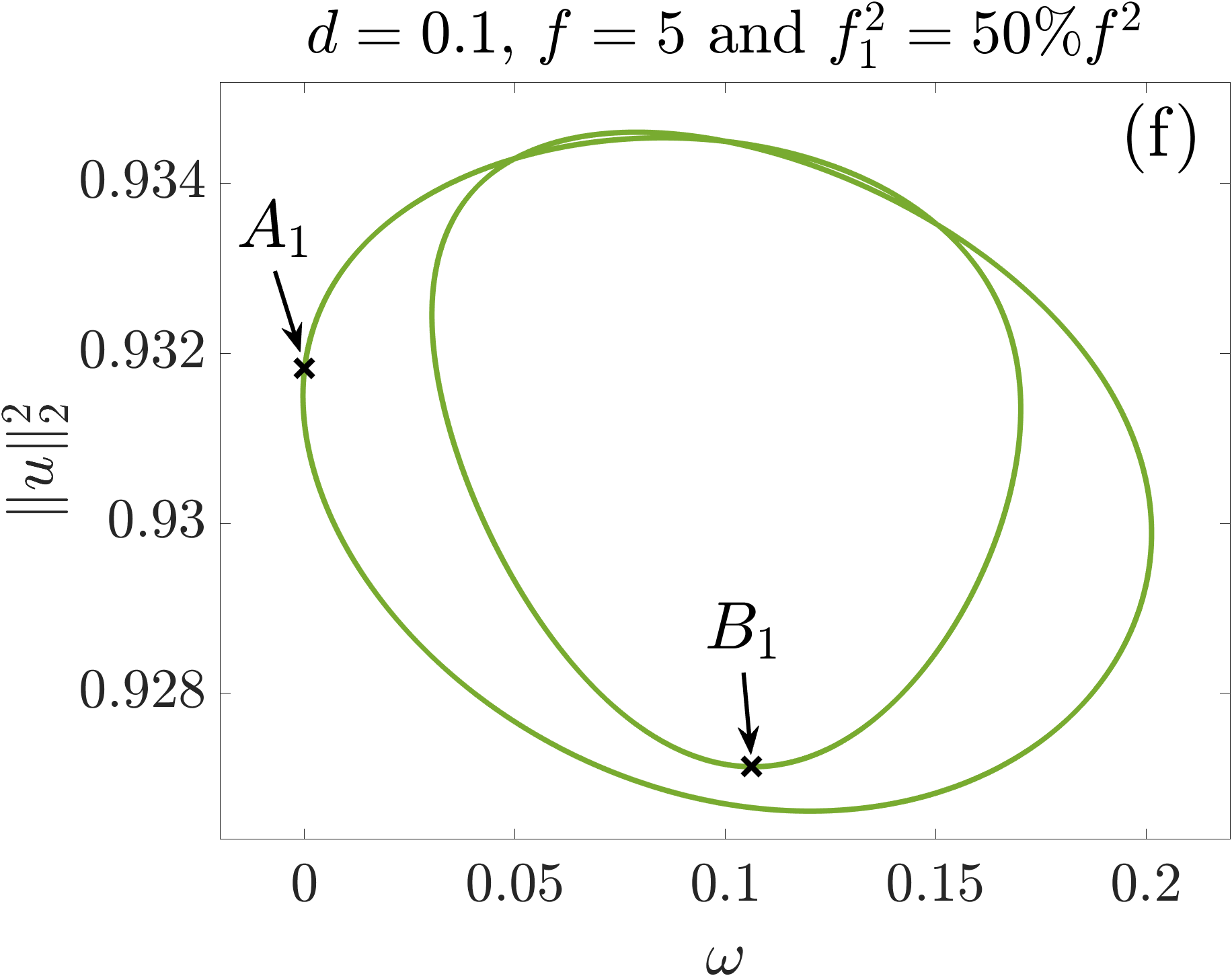} \\[0.5cm]
\includegraphics[width=\columnwidth]{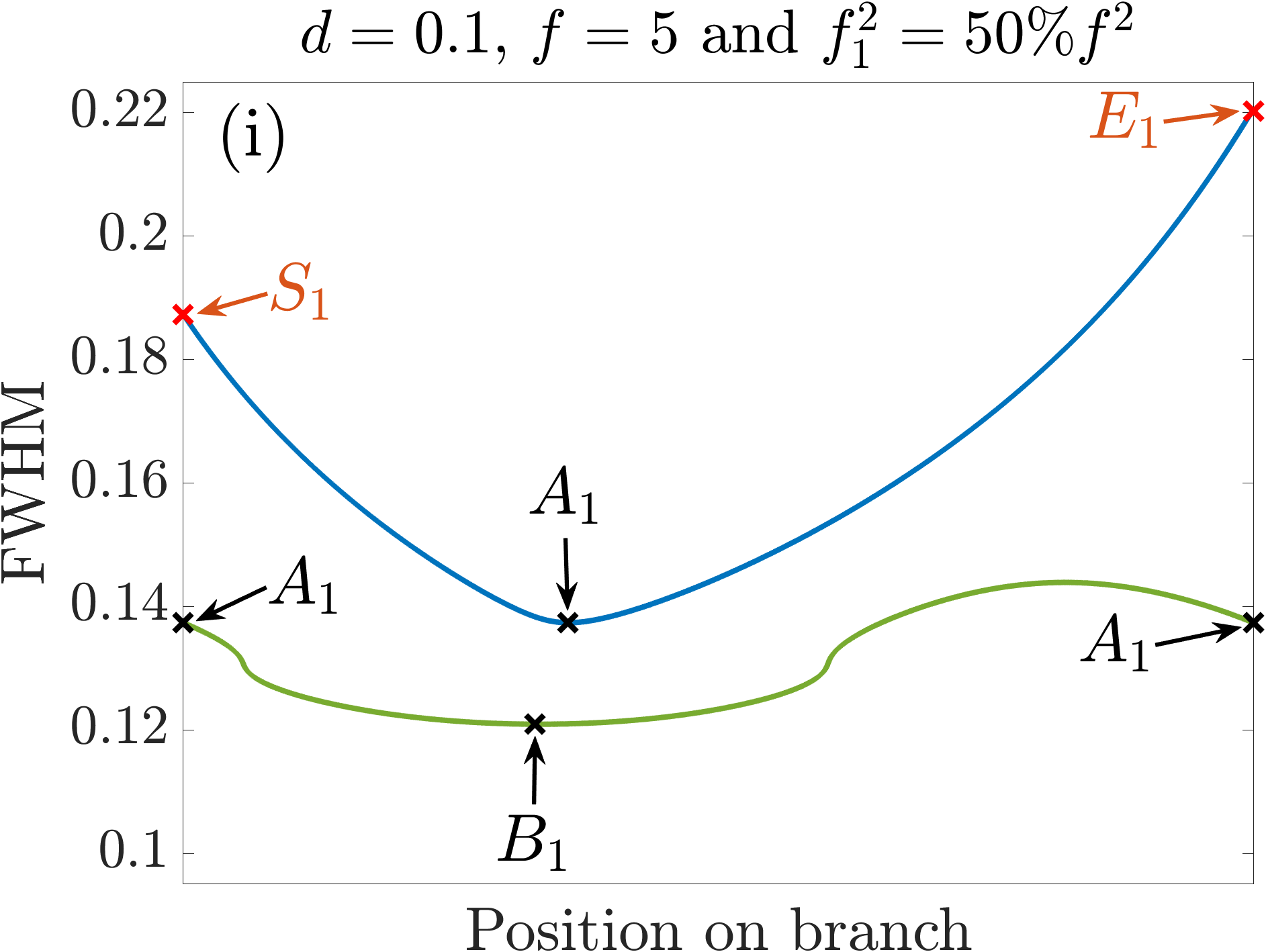}
\end{minipage}
\caption{Branches show intracavity power $\|u\|_2^2=\frac{1}{2\pi}\int_0^{2\pi} |u(s)|^2 \, ds$ of the soliton $u$ plotted vs. values of $\zeta_0$ or $\omega$. First row: blue branch as achieved by the first $\zeta_0$-optimization. Second row: green branch as achieved by the first $\omega$-optimization. Third row: FWHM along these branches. Columns correspond to different values of $d, f_1, f$.}
\label{fig:Branches}
\end{figure*}

In Fig.~\ref{fig:Branches} we have illustrated Step 2 and 3 for three different values of the parameters $d$, $f$ and $f_1$. In the first row Fig.~\ref{fig:Branches}(a)-(c) we are plotting $\zeta_0$-branches, i.e., the intracavity power of the soliton, given by $\|u\|_2^2=\frac{1}{2\pi}\int_0^{2\pi} |u(s)|^2 \, ds$,  versus the value of $\zeta_0$ (Step 2, $\zeta_0$-optimization). The points $A_1$ indicate the optimal soliton with the smallest FWHM and they are located near a turning point. They serve as starting points for the subsequent $\omega$-continuation (Step 3). The $\omega$-branches, i.e., intracavity power of the soliton versus the value of $\omega$, depicted in Fig.~\ref{fig:Branches}(d)-(f) turn out to be closed loops. The points $B_1$ indicate the optimal soliton on the closed $\omega$-loop. In the last row Fig.~\ref{fig:Branches}(g)-(i) we illustrate the optimality of the points $A_1$, $B_1$ by plotting the value of FWHM along the $\zeta_0$-branches (blue) and the $\omega$-loops (green). The FWHM is depicted as a function of normalized arc length of the corresponding curves. Since the $\zeta_0$-curves are unbounded, we decided to plot the FWHM between the reference points $S_1$ (start) before the relevant $1$-solitons begin and $E_1$ (end) after the relevant $1$-solitons have been passed. 

\medskip
 
An iteration of the $\zeta_0$- and $\omega$-optimization steps (until a desired accuracy is reached) provides similar pictures. In our numerical experiments we always performed three optimizations in both of the variables $\zeta_0$ and $\omega$ (unless stated otherwise).

\makeatletter\onecolumngrid@push\makeatother
\makeatletter\onecolumngrid@pop\makeatother

\section*{Appendix C: Heuristic for finding localized solitons in the case of pumping two arbitrarily distanced modes} \label{Heuristic_arbitrary_k1}

\noindent By considering additional bifurcations we will demonstrate how the heuristic from Section~\ref{Heuristic} can be adapted to arbitrary values of $k_1 \geq 2$. A first observation is that the very same heuristic as used in Section~\ref{Heuristic} would lead to solitons which are not only $2\pi$- but in fact $2\pi/k_1$-periodic, i.e., the algorithm detects no 1-solitons. This is essentially due to the fact that starting from a constant solution any kind of parameter-continuation will develop solutions that have the shape of the pump.

\medskip

\begin{figure*}
\centering
\begin{minipage}[t]{0.3\textwidth}
\includegraphics[width=\columnwidth]{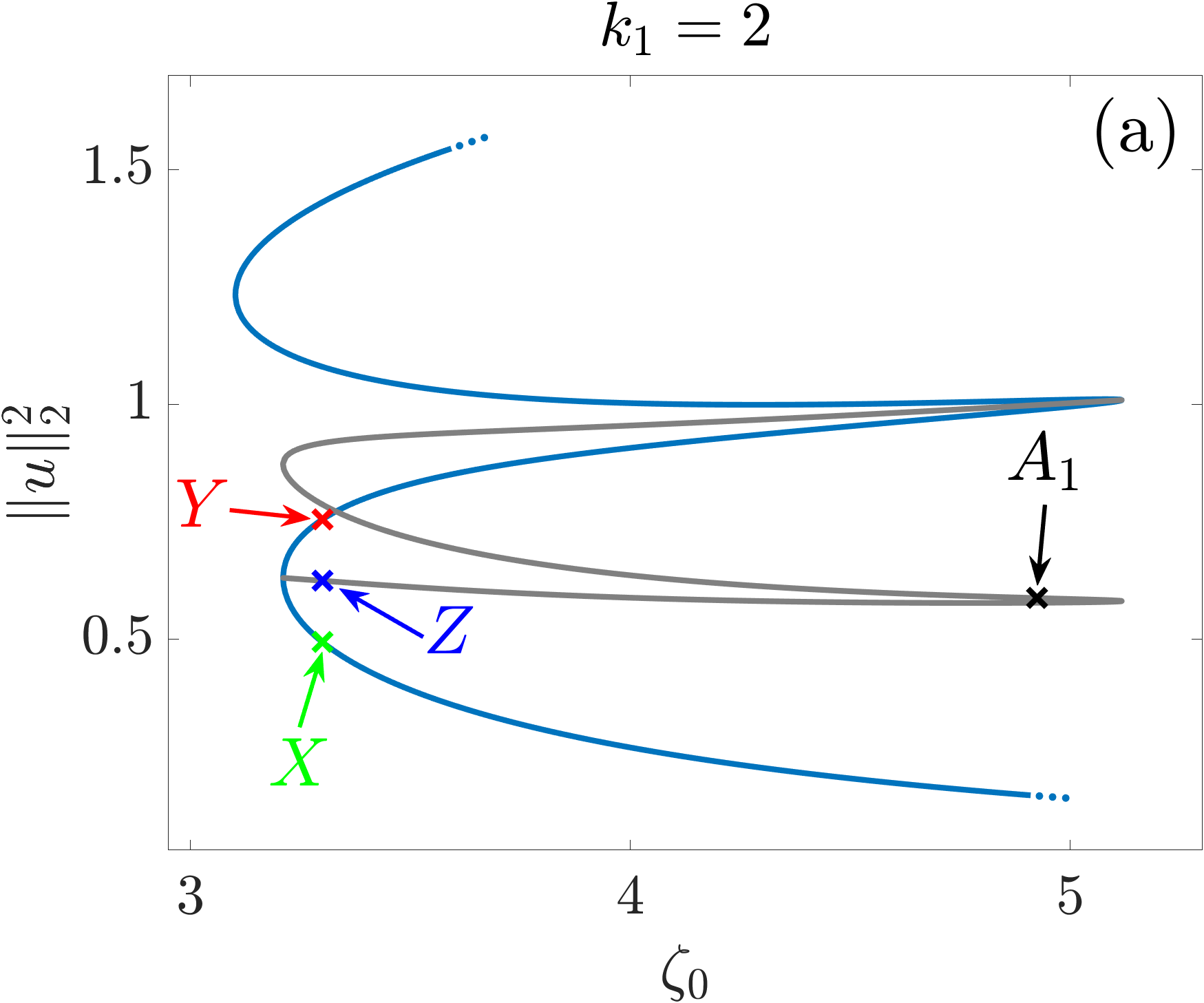} \\[0.5cm]
\includegraphics[width=\columnwidth]{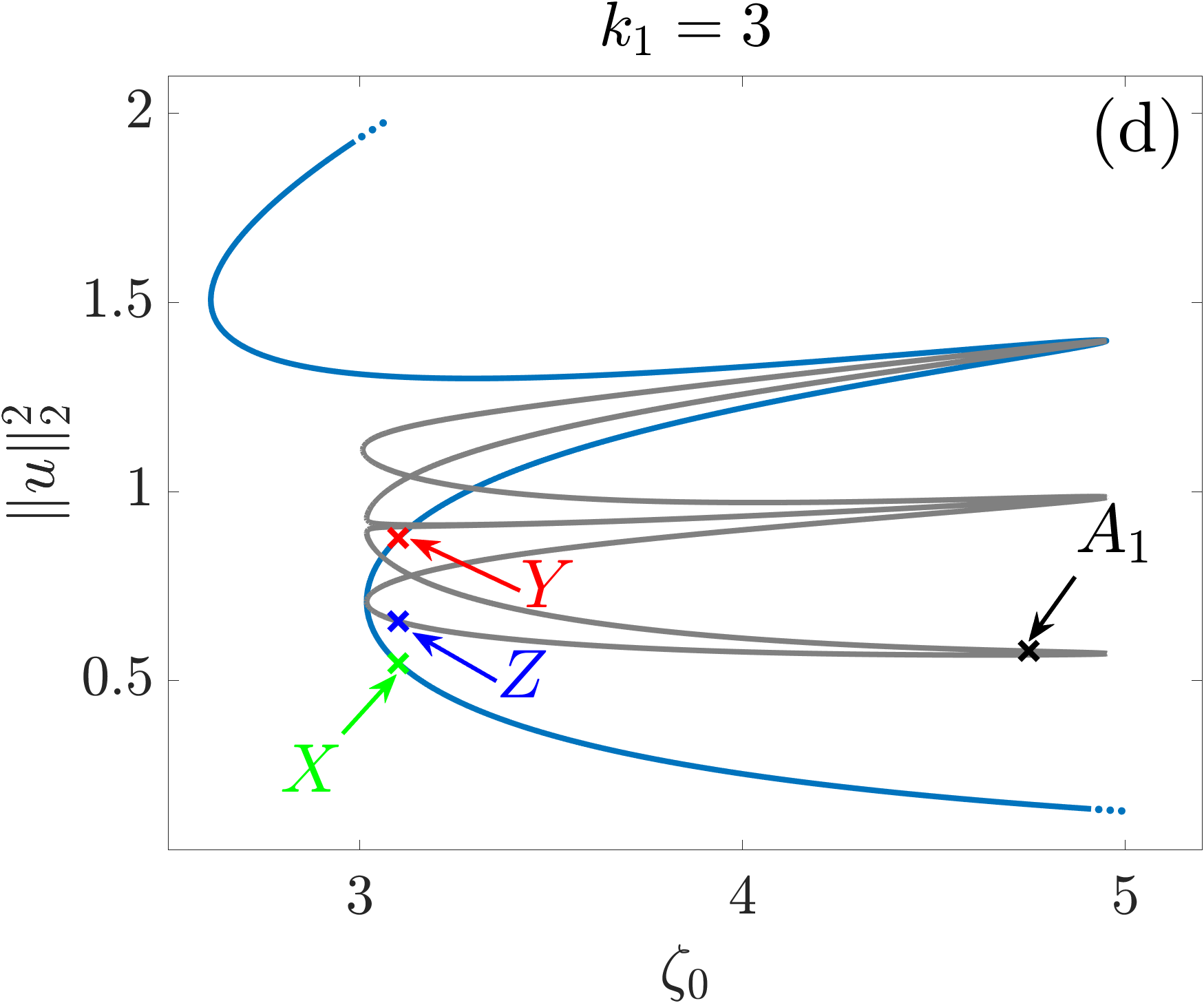}
\end{minipage} \hspace{0.5cm}
\begin{minipage}[t]{0.305\textwidth}
\includegraphics[width=\columnwidth]{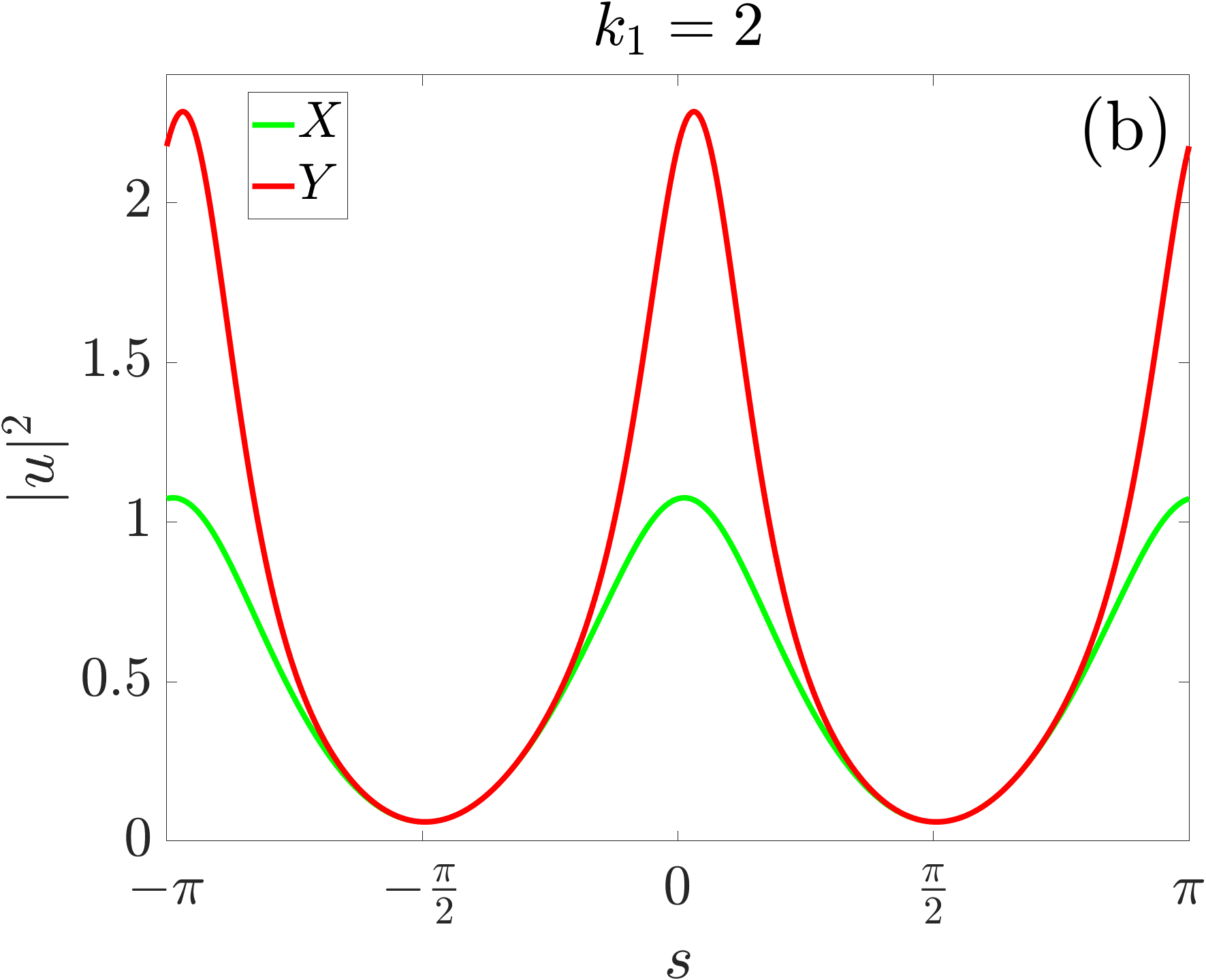} \\[0.5cm]
\includegraphics[width=\columnwidth]{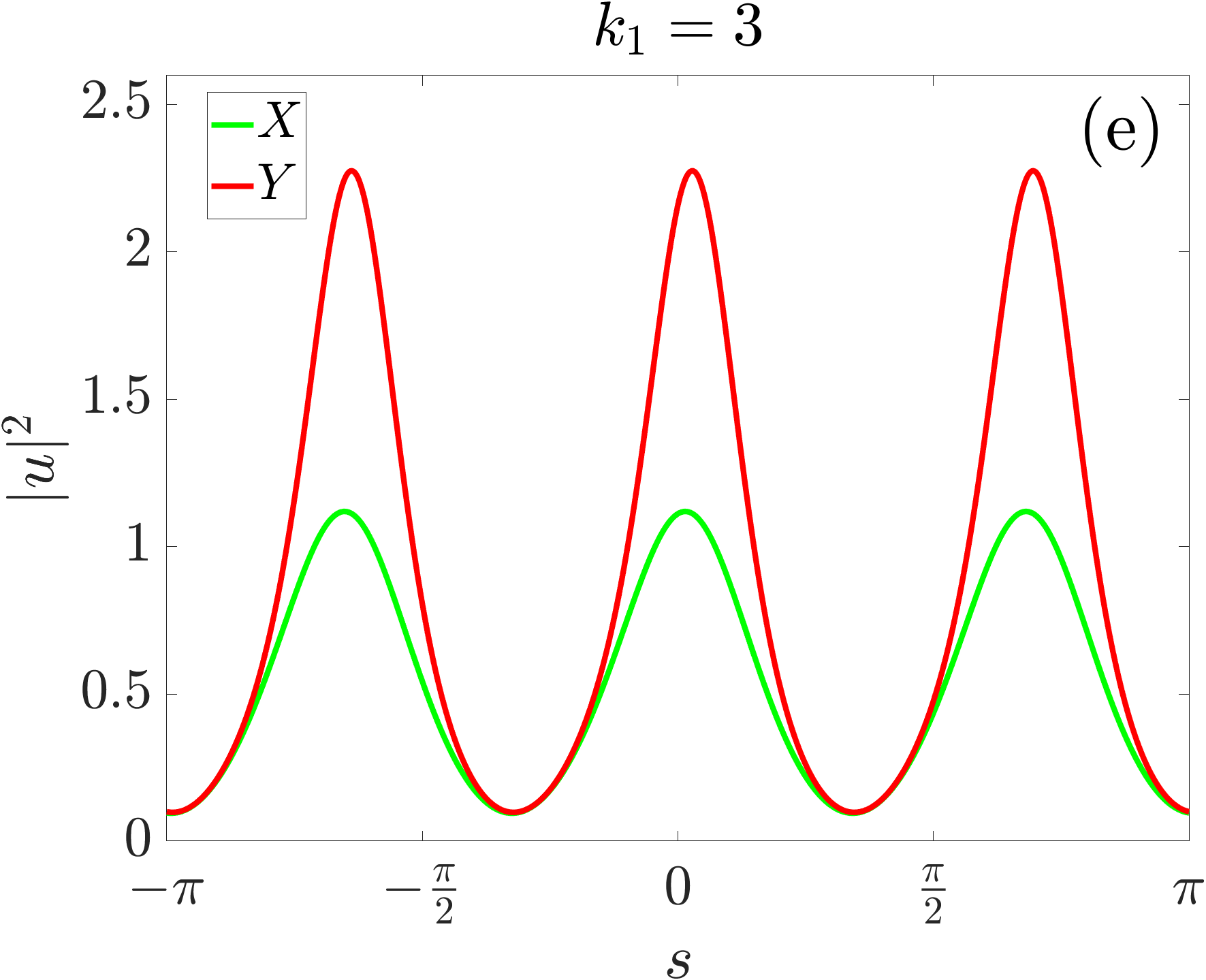}
\end{minipage} \hspace{0.5cm}
\begin{minipage}[t]{0.305\textwidth}
\includegraphics[width=\columnwidth]{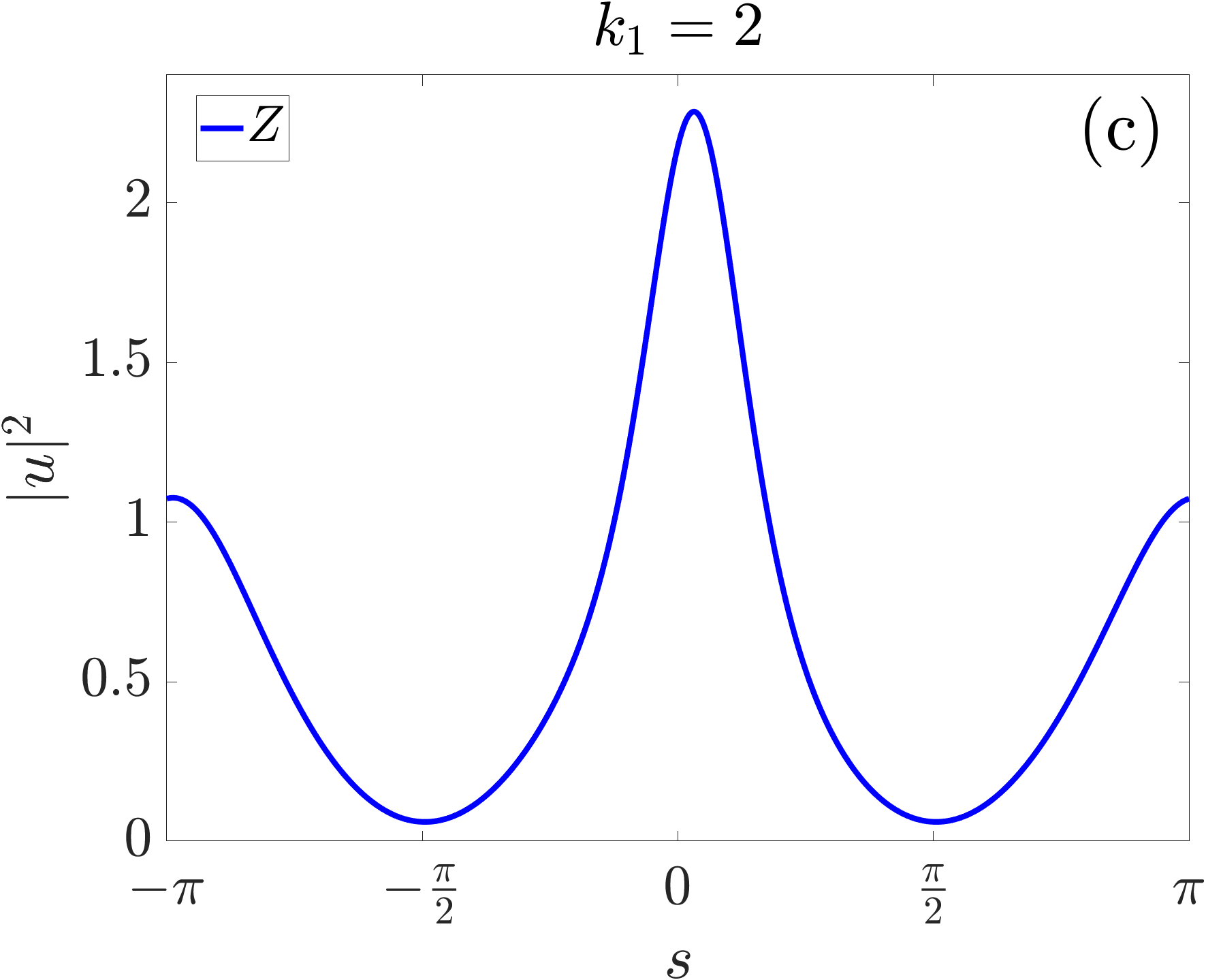} \\[0.5cm]
\includegraphics[width=\columnwidth]{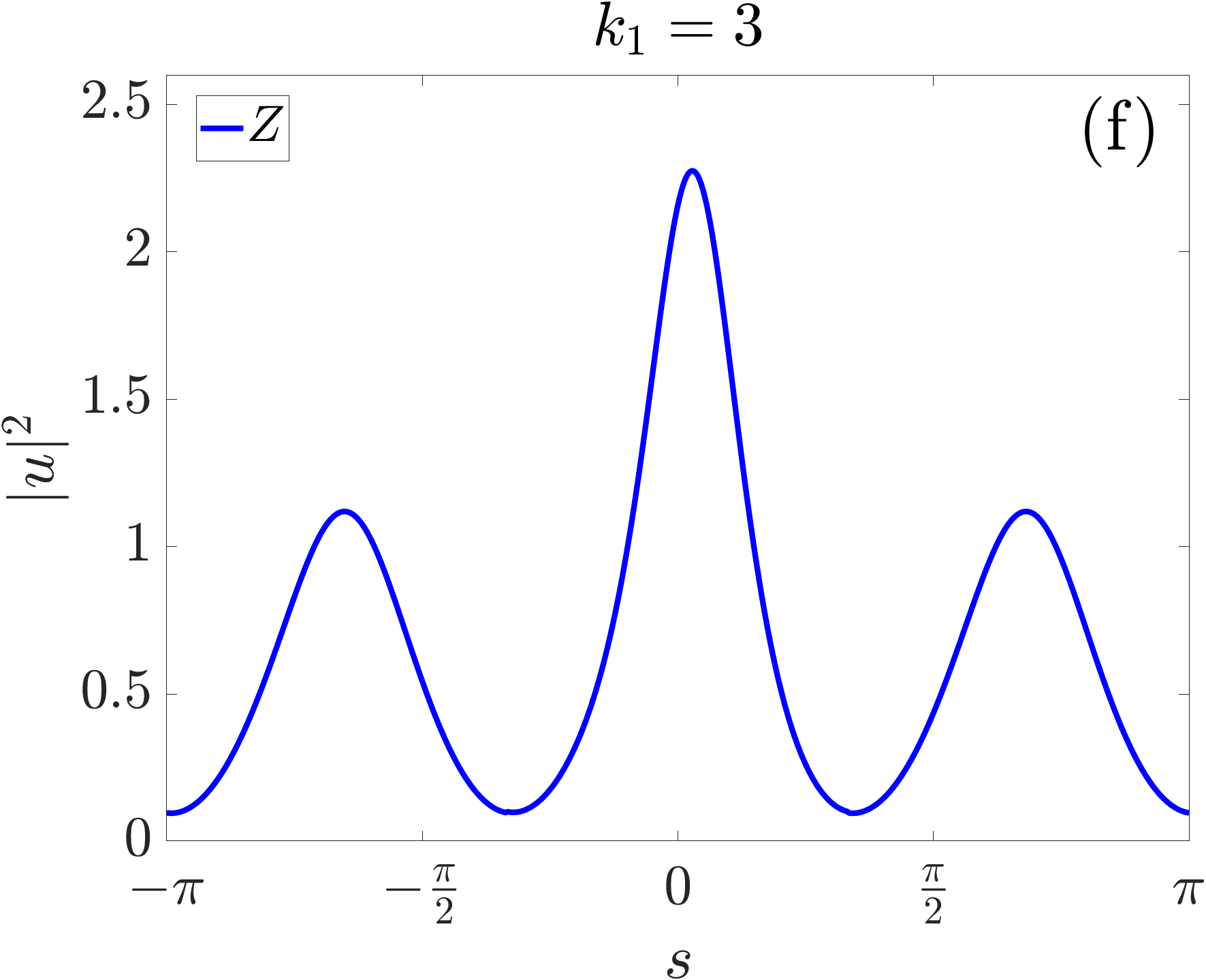}
\end{minipage}
\caption{Example for $d=0.1$, $f=2$ and $f_1^2=25\%f^2$. First column: branches show intracavity power $\|u\|_2^2=\frac{1}{2\pi}\int_0^{2\pi} |u(s)|^2 \, ds$ of the soliton $u$ plotted vs. $\zeta_0$. Blue branch as achieved by first $\zeta_0$-continuation and gray branch obtained from first bifurcation from blue branch. Second and third column: spatial power distribution of solutions used for branch-switching. Plots (a)-(c) correspond to the case $k_1=2$ while plots (d)-(f) correspond to $k_1=3$.}
\label{fig:Glueing}
\end{figure*}

However, in contrast to the case $k_1=1$, we also detect bifurcations this time. The idea of the adapted heuristic is to switch in every $\zeta_0$-optimization step to a bifurcating branch containing 1-solitons. For $d=0.1$, $f=2$, $f_1^2=25\%f^2$ this is illustrated in Fig.~\ref{fig:Glueing}(a),(d) for $k_1=2,3$. The gray branch is the new additional branch bifurcating from the first continued (blue) branch in $\zeta_0$ and $A_1$ indicates the optimal point with the minimal FWHM on that branch. The point $A_1$ is then used as starting point for the subsequent $\omega$-continuation and from here on we can once again iterate the whole process. 

\medskip

The mentioned bifurcations turn out to be not of simple nature in general. For example, if $k_1$ is odd, \texttt{pde2path} detects no bifurcations at all (which may be due to an even number of eigenvalues crossing zero simultaneously). However, we can easily overcome this issue by using an interpolation trick for branch-switching. For that, we consider a $\zeta_0$-value near a turning point, where we find two distinct solutions (named $X$ and $Y$) for one and the same value of $\zeta_0$. In Fig.~\ref{fig:Glueing}(a) we used $\zeta_0=3.3$ and in Fig.~\ref{fig:Glueing}(d) we used $\zeta_0=3.1$ for this purpose and marked the mentioned solutions in red and green, respectively. Fig.~\ref{fig:Glueing}(b) and Fig.~\ref{fig:Glueing}(e) show the spatial power distributions of $X$ and $Y$. It turns out that a 1-soliton-like state, which is not $2 \pi/k_1$-periodic anymore, can be glued together from parts of these solutions. The resulting soliton $Z$ is marked in blue in Fig.~\ref{fig:Glueing}(a) and Fig.~\ref{fig:Glueing}(d) and its spatial power distribution is given in Fig.~\ref{fig:Glueing}(c) and Fig.~\ref{fig:Glueing}(f). The interpolated soliton serves as starting point for another $\zeta_0$-continuation yielding the gray branch which actually is a branch which bifurcates from the original curve and connects two of its turning points.

\makeatletter\onecolumngrid@push\makeatother
\makeatletter\onecolumngrid@pop\makeatother

\bibliographystyle{apsrev4-1}
\bibliography{bibliography}

\end{document}